 \documentclass[times,twocolumn,final]{elsarticle}

\usepackage{jasr}
\usepackage{framed,multirow}
\usepackage{graphicx}

\providecommand{\linenumbers}{}
\providecommand{\nolinenumbers}{}

\usepackage{amssymb}
\usepackage{latexsym}

\usepackage{url}
\usepackage{xcolor}
\definecolor{newcolor}{rgb}{.8,.349,.1}

\usepackage[citebordercolor=white]{hyperref}

\journal{Advances in Space Research}
\begin{document}

\verso{Open clusters \textit{NGC 2266 and NGC 2324}}

\begin{frontmatter}

\title{Stellar Dynamics and Evolution of the Intermediate-Age Open Clusters NGC 2266 and NGC 2324}

\author[1]{Cyrus  Raj}

\author[2]{D. Bisht}

\author[2,3]{Ashish  Raj}

\cortext[cor1]{Authors Email:\\
  \makebox[2.2cm][l]{} cyrusraj1996@gmail.com \\
  \makebox[2.2cm][l]{} devendrabisht297@gmail.com (Corresponding Author) \\
  \makebox[2.2cm][l]{} ashishpink@gmail.com
}

\affiliation[1]{{organization}=Independent Researcher, city={Panipat},   postcode={132106} , state={Haryana},  country={India}}
\affiliation[2]{organization={Indian Centre for Space Physics}, addressline={466 Barakhola, Netai Nagar}, city={Kolkata},   postcode={700099}, state={West Bengal}, country={India} }
\affiliation[3]{organization={Uttar Pradesh State Institute of Forensic Science (UPSIFS)}, addressline={Aurawan, P.O. Banthra}, city={Lucknow}, postcode={226401},    state={(U.P)},       country={India}}


\begin{abstract}
We present a refined astrometric and photometric analysis of the well-studied intermediate-age open clusters NGC 2266 and NGC 2324 using high-precision Gaia DR3 data, complemented by 2MASS and LAMOST DR7 catalogs. Probable cluster members are identified using unsupervised machine learning techniques. We apply both Gaussian Mixture Models (GMMs) and \texttt{pyUPMASK}. We find that the GMM-based membership sample yields a cleaner, more coherent cluster sequence in the Gaia CMDs than pyUPMASK. We identified 719 and 852 high-probability members ($P \geq 0.7$) for NGC 2266 and NGC 2324, respectively. Using the parallax method, we determine distances of 3.55 $\pm$ 0.23~kpc for NGC 2266 and 4.18 $\pm$ 0.24~kpc for NGC 2324. The radius estimates for both clusters are 7.23 $\pm$ 0.47 pc and 10.94 $\pm$ 0.63 pc. Isochrone fitting estimated ages of $1.1 \pm 0.1$~Gyr for NGC 2266 and $790 \pm 150$~Myr for NGC 2324. These age estimates were derived assuming metallicities of $Z = 0.0084$ and $Z = 0.0038$, respectively. The King profile fitting indicates that both clusters exhibit compact, well-defined radial structures. Their tidal radii are $8.84'$ (9.13 pc) for NGC 2266 and $10.97'$ (13.34 pc) for NGC 2324. The slopes of the present-day mass functions are $1.13\pm0.18$ for NGC 2266 and $1.24\pm0.19$ for NGC 2324, indicating a deficiency of low-mass stars. The derived mass-function slopes are consistent with dynamical evolution in both clusters. The clusters exhibit short relaxation times, while only NGC 2324 shows a mild indication of mass segregation. This study highlights the power of Gaia astrometry to resolve internal structures within open clusters and refine their dynamical parameters.
\end{abstract}


\end{frontmatter}



\linenumbers
\section{Introduction}\label{sec1}

Open clusters (OCs), also known as galactic clusters, are loosely bound groups of stars formed within the same molecular cloud. They originate through gravitational collapse and fragmentation of gas and are typically found near the Galactic plane. This area hosts many complex physical processes within these giant molecular clouds. During star formation, only part of the gas forms stars, while the rest remains as leftover material in the cluster environment. These stars share similar ages, chemical compositions, and kinematic properties. This makes OCs excellent laboratories for studying stellar evolution and the structure of the Galactic disk. Using OCs provides detailed statistics and insights into their dynamics and transformations. Intermediate OCs help to deepen our understanding of stellar evolutionary phases. The present-day mass function (PDMF) represents the distribution of stellar masses and serves as a key diagnostic for internal cluster dynamics. Variations in the PDMF, such as mass segregation, suggest that massive stars may migrate to the cluster center, whereas lower-mass stars move outward. Star clusters evolve through internal processes, such as stellar winds, interactions, and mass loss, as well as external forces, including encounters with molecular clouds, supernovae, and the Galactic tidal field. Determining cluster parameters, such as age, distance, reddening, and mass, is often complicated by field-star contamination. OCs are fundamental laboratories for understanding stellar evolution, Galactic disk structure, and stellar dynamics, since their member stars share similar ages, distances, and chemical compositions (Janes \& Adler, 1982; Friel, 1995; Piskunov et al., 2006; Cantat-Gaudin et al., 2020). Their structural and dynamical properties provide important constraints on star formation, internal dynamical evolution, and interaction with the Galactic tidal field (Portegies Zwart et al., 2010; Sagar \& Griffiths, 1998).
\\
To address this, various membership estimation methods have been developed over the past several decades. In this paper, we use two unsupervised clustering algorithms: the Gaussian Mixture Model (GMM) and py\textit{UPMASK}, to identify probable member stars with high precision and accuracy. With \textit{Gaia} DR3 \citep{bib8}, we can access stars within 21 mag. To ensure reliable parameter estimation, we limited our analysis to stars brighter than G = 20 mag, where Gaia DR3 provides high completeness and reliable astrometric measurements. At fainter magnitudes, the uncertainties in parallax and proper motion increase significantly, reducing the reliability of membership determination and derived physical parameters (Gaia Collaboration et al., 2023; Riello et al., 2021). Therefore, the completeness limit of G = 20 mag was adopted for membership selection and subsequent analyses, including luminosity function, mass function, and dynamical parameter estimation. Dynamical mass segregation and stellar evaporation contribute to the preferential loss of low-mass stars, thereby steepening the mass function slope. As a result, the PDMF gradually flattens over time, as discussed in later sections of this work and \citep{bib34, bib19, bib24}.
In this study, we focus on the OCs NGC 2266 and NGC 2324, located in the constellations Gemini and Monoceros, respectively. Both clusters lie in the third Galactic quadrant, a region known for significant field star contamination. These clusters had been the subject of numerous studies using photometry, spectroscopy, Gaia-based membership, and blue straggler identification. The popular catalog \citep{bib43} used pyUPMASK \citep{bib43} to compute membership probabilities, while the latest catalog \citep{bib216} employed DBCSCAN to derive mean parameters, including radius, distance, and galactocentric coordinates. Our primary aim is to precisely derive fundamental parameters (e.g., reddening, metallicity, age) and masses, relaxation times, and other evolution-related parameters for both clusters. We used only probable member stars identified through astrometric filtering. Here is the available information for both clusters in the literature-\\ 
\\
\textbf{NGC 2266} ($\alpha_{J2000}$=06h43m19s, $\delta_{J2000}$=+26$^\circ$58', $l=187.8^\circ$, $b=10.3^\circ$; \citep{bib55}) is a compact star cluster in the Gemini constellation, known to have many stars in the main sequence. It is particularly noted for its 11 red giants and two blue straggler stars \citep{bib112}. A set of 12 binary stars has also been detected \citep{bib112}. This open cluster has been of interest to various scientists since the 1990s \citep{bib112, bib116, bib55, bib57, bib117, bib56, bib54}. Most blue stars in the cluster core are hydrogen-burning stars. However, many stars have progressed beyond this evolutionary stage because their central hydrogen has been depleted. They now burn hydrogen in a shell around the stellar nucleus, appearing as bright red giants. The brightest stars in the image are foreground stars, which do not belong to the cluster population.

\textbf{NGC 2324} ($\alpha$$_J$$_2$$_0$$_0$$_0$=07h04m02, $\delta$$_J$$_2$$_0$$_0$$_0$=+01°04', l= 213$^o$.447, b= 3$^o$.297; \citep{bib54}) is a distant rich star cluster from the Monoceros constellation known to be located at ~35$^o$ from the anticenter beyond Perseus spiral arm situated close to the celestial equator, first discovered by Cuffey (1941). The cluster has a history of studies of red giants \citep{bib58} and contains binaries. NGC 2324, also known as Cr 125 (Collinder 1931), was classified as a relatively young, metal-deficient open cluster based on high-resolution data and CCD VI$_K$$_C$ and CT$_1$ photometry \citep{bib2}. This OC has been studied since 1992 by \citep{bib118, bib60, bib2, bib59, bib55, bib41,bib54}.\\
\\

Although NGC 2266 and NGC 2324 have been extensively studied in the Gaia era, including membership analyses based on DR2 (e.g., Cantat-Gaudin et al. 2020) and recent DR3-based catalogs (e.g., Hunt \& Reffert 2023), several aspects remain open for refinement. First, most previous large-scale catalogs focused on homogeneous parameter compilation rather than on detailed structural and dynamical reanalysis using cluster-specific optimization. Second, earlier Gaia-based studies typically used either a single clustering technique or automated density-based methods, without comparing multiple probabilistic approaches within the same dataset. 

In this work, we perform a dedicated reanalysis of Gaia DR3 data, with photometry extending to $G = 20$ mag, and conduct a comparative assessment of two independent unsupervised membership techniques (GMM and pyUPMASK). This enables us to (i) evaluate the robustness of different membership approaches in crowded Galactic fields, (ii) obtain a clean and internally consistent membership sample for subsequent analysis, and (iii) derive updated structural and dynamical parameters for both clusters. The adopted membership sample provides the basis for determining the present-day mass function, mass-segregation signatures, and relaxation times.

The paper is organized as follows. Section \ref{2} describes the data used in the analyses. The section \ref{4} describes the mean proper motion, membership determination, and comparison between the GMM and pyUPMASK methods. The fundamental structural parameters, including the evaluation of distances using mean parallax and radial density profiles using the King model, will be discussed in Sections \ref{sec5} and \ref{rdp}. Then, a photometric study examines the fundamental physical parameters of interstellar reddening, metallicity, age, and distance, as well as the detection of blue straggler stars, in the subsections that follow in Section \ref{7}. The dynamical evolution can be examined by interpreting the luminosity function, mass function, mass segregation, relaxation time, and dissociation time in the subsections of Section \ref{8}. We present the results and the discussion in Section \ref{9}.

\begin{figure*}
\begin{center}
  \begin{minipage}[b]{0.46\textwidth}
    \includegraphics[width=\textwidth]{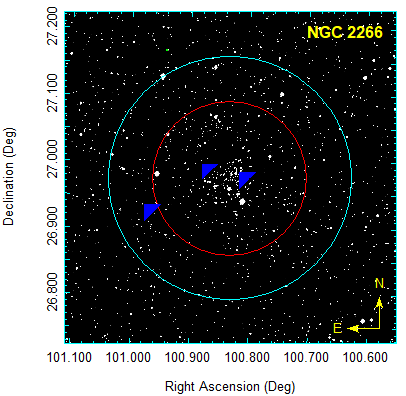}
  \end{minipage}
  \hfill
  \begin{minipage}[b]{0.447\textwidth}
    \includegraphics[width=\textwidth]{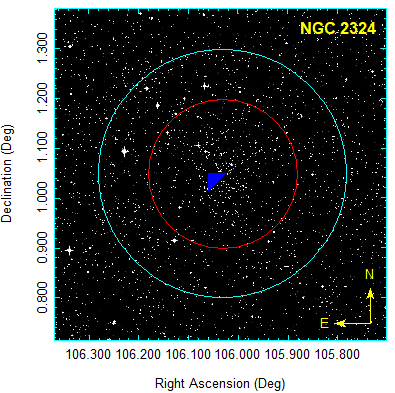}
  \end{minipage}
\caption{Identification maps of NGC~2266 and NGC~2324 obtained from the Digitized Sky Survey. The red inner circles represent the cluster radii ($7'$ for NGC~2266 and $9'$ for NGC~2324), while the outer circles in sky color denote the field of analysis ($11' \times 11'$ and $15' \times 15'$ respectively), based on \textit{Gaia}~DR3 data. The blue triangles in the left figure represent identified BSS. }
\label{Figure.1}
\end{center}
\end{figure*}

\section{Data} \label{2}
\begin{figure*}[h!]
\centering
   \begin{minipage}[b]{0.32\textwidth}
    \includegraphics[width=\textwidth]{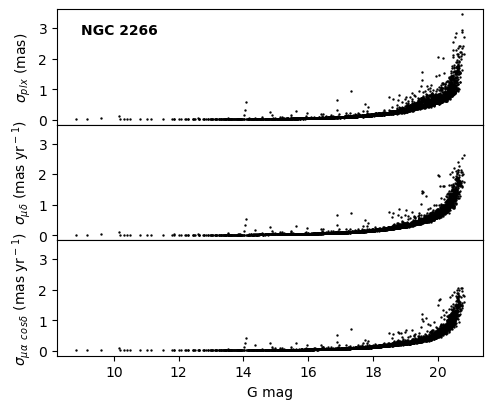}
  \end{minipage}
  \hfill
  \begin{minipage}[b]{0.33\textwidth}
    \includegraphics[width=\textwidth]{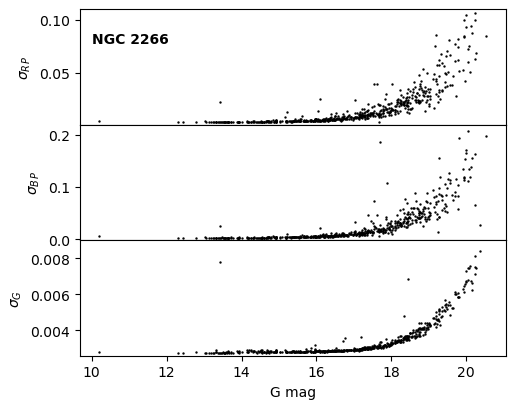}
  \end{minipage}
  \hfill
  \begin{minipage}[b]{0.32\textwidth}
    \includegraphics[width=\textwidth]{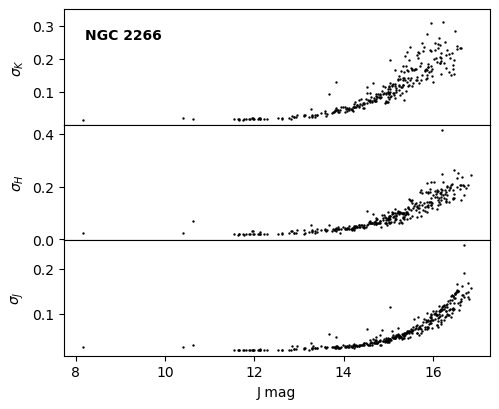}
   \end{minipage}
\hfill
  \begin{minipage}[b]{0.32\textwidth}
    \includegraphics[width=\textwidth]{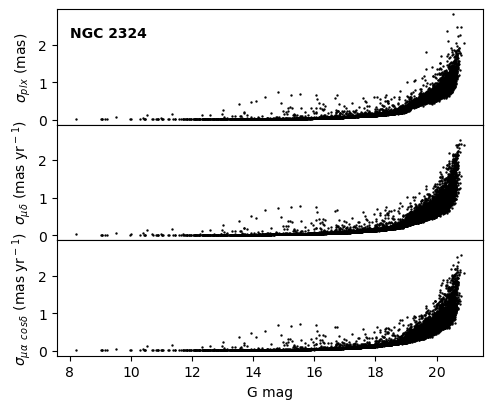}
  \end{minipage}
  \hfill
  \begin{minipage}[b]{0.33\textwidth}
    \includegraphics[width=\textwidth]{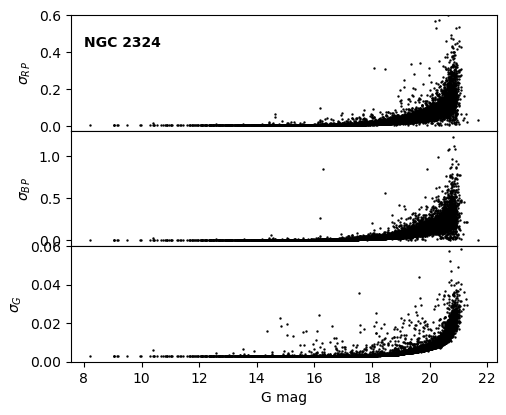}
  \end{minipage}
  \hfill
  \begin{minipage}[b]{0.32\textwidth}
    \includegraphics[width=\textwidth]{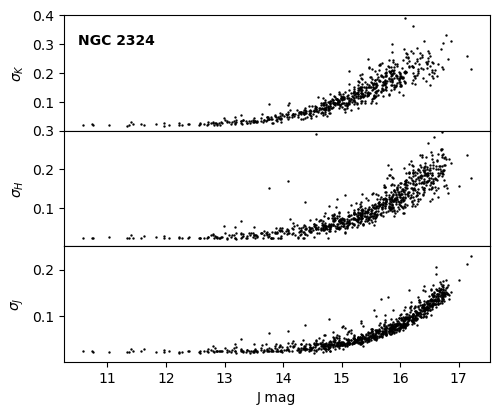}
   \end{minipage}
   \caption{(a) Errors in \textit{Gaia} proper motions and parallax as a function of $G$ magnitude. (b) Photometric uncertainties in \textit{Gaia} bands ($G$, $G_{\rm BP}$, $G_{\rm RP}$). (c) Errors in 2MASS photometry ($J$, $H$, $K$) vs. $J$ magnitude.}
\label{fig2}
\end{figure*}

The clusters were homogeneously analyzed using archived data from \textit{Gaia} DR3, 2MASS, and LAMOST surveys in the target field size ($11' \times 11'$) for NGC 2266 and ($15' \times 15'$) for NGC 2324 (see Figure\ref{Figure.1}). These field sizes extend well beyond the reported limiting and tidal radii of the clusters, ensuring complete spatial coverage of the cluster regions and sufficient surrounding area for reliable estimation of the background stellar density. This approach is essential for accurate membership determination and structural analysis, including King profile fitting, while minimizing the effects of field-star contamination. The \textit{Gaia} catalog contains high-precision parameters for 1.8 billion celestial objects. The \textit{Gaia} Data Release 3 \citep{bib8} provides a complete dataset comprising five astrometric parameters (positions, parallaxes, proper motions) and three photometric band parameters ($G$, $BP$, $RP$ magnitudes), enabling us to investigate OCs with high accuracy. The high-precision astrometric measurements provided by Gaia DR3 have revolutionized studies of stellar populations, enabling robust determinations of stellar parameters, kinematics, and Galactic origins \citep{deniz25}. Under \textit{Gaia} DR3, we obtained 3,461 stars for NGC 2266 and 11,857 for NGC 2324. Many stars in the selected fields are either intrinsically faint or at large distances, leading to increased uncertainties in their astrometric parameters, such as parallax and proper motion. These uncertainties increase markedly at fainter magnitudes, thereby reducing the reliability of distance estimates and membership determination. Therefore, we evaluated the dependence of astrometric and photometric uncertainties on magnitude, as shown in Figure \ref{fig2}, and restricted our analysis to stars brighter than the completeness limit (G $\le$ 20 mag) to ensure reliable parameter estimation. Photometric distances and cluster parameters were determined using Gaia G magnitudes and $BP-RP$ colors, combined with isochrone fitting and parallax-based distance estimates. To learn more about the errors in these parameters, we plotted the errors in proper motions (PM) and parallax with $G$ magnitude along with photometric bands with magnitudes $ J$, $H$, and $ K$ vs $J$ and $ G$, $BP-RP$ vs $G$  in (Fig. \ref{fig2}). The value of the mean PM error for the studied stars is 0.01 mas/yr for $G$ mag up to 20 mag. The mean error in parallax is $\sim 0.28$ mas for stars brighter than 20 $G$ mag. The median errors in $G$, $BP$, and $RP$ are $\sim 0.003$, $\sim 0.03$, and $\sim 0.02$ for $G \leq 20$ mag. \\\\
The Two Micron All Sky Survey (2MASS), commonly known as 2MASS \citep{bib100}, provides stellar parameters for approximately 470 million objects. Its near-infrared magnitudes $J$, $H$, and $K$ were used to determine one of the photometric parameters of the interstellar reddening of OCs in subsection \ref{redenn}. 
The LAMOST DR7 catalog (Data Release 7) had accumulated more than 21 million spectra by the end of 2022, comprising 11 million low-resolution spectra and 10 million medium-resolution spectra, as mentioned in \citep{bib98}. In our study, we used parameters from the LAMOST DR7 LRS, which were determined by the LAMOST Stellar Parameter Pipeline \citep{bib99}, including effective temperature, surface gravity, metallicity, and radial velocity. The ESO Digitized Sky Survey (DSS) is a web-based application that provides online access to the digitized sky and is used to extract random sky regions from the DSS server. The Space Telescope Science Institute (STScI) has taken these survey images through their Guide Star Survey group, shown in (Figure \ref{Figure.1}). These batch files are available in DSS1 and DSS2; we have used the DSS1 mode. 
The \citep{bib43} has memberships only up to 18 magnitudes. With higher-resolution data from \citep{bib8}, we obtained memberships up to 20 magnitudes in the G band using two membership methods.\\\\

\vspace{-1cm}
\section{Mean Proper Motion and Cluster Membership Determination} \label{4}

\begin{figure*}[h!]
   \begin{minipage}[b]{0.5\textwidth}
    \includegraphics[width=\textwidth]{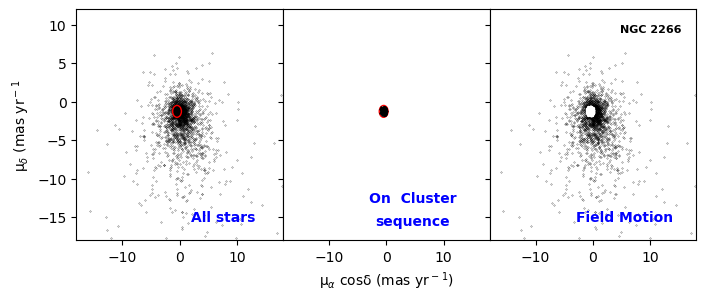}
  \end{minipage}
  \hfill
  \begin{minipage}[b]{0.5\textwidth}
    \includegraphics[width=\textwidth]{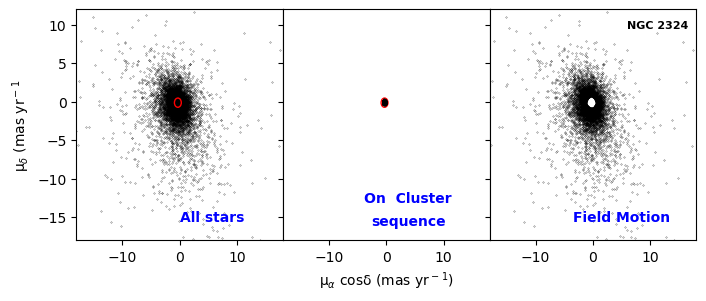}
  \end{minipage}
  \label{a}
\hfill
  \begin{minipage}[b]{0.48\textwidth}
    \includegraphics[width=\textwidth]{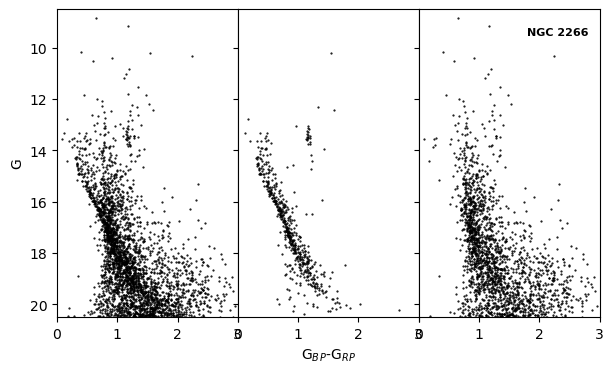}
   \end{minipage}
   \hfill
  \begin{minipage}[b]{0.48\textwidth}
    \includegraphics[width=\textwidth]{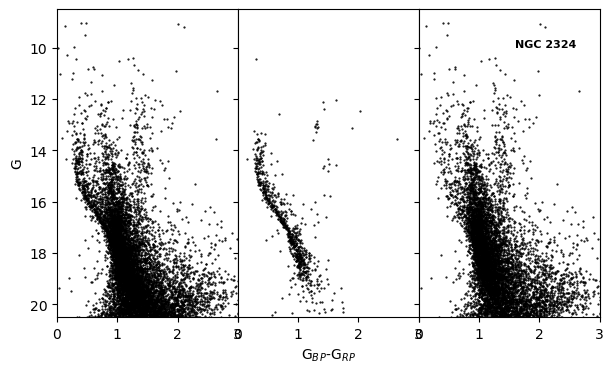}
  \end{minipage}
\caption{Top panels: Vector Point Diagrams (VPDs) showing proper motion in right ascension $\mu_{\alpha} \cos \delta$ versus proper motion in declination ($\mu_{\delta}$) for stars in the cluster regions of NGC 2266 (left) and NGC 2324 (right). Cluster members are concentrated around the mean proper motion values, while field stars are more widely distributed. Bottom panels: Gaia color–magnitude diagrams (CMDs) plotted as G magnitude versus $(G_{BP}-G_{RP})$ color. The left panels show all stars in the selected field, the middle panels show probable cluster members, and the right panels show field stars rejected from the membership selection.}
    \label{fig3}
\end{figure*}

Cluster members are expected to exhibit proper motions clustered around a common mean value due to their shared kinematic properties. This concentration can be visualized using Vector Point Diagrams (VPDs), which display the proper motion components ($\mu_{\alpha}\cos\delta$, $\mu_{\delta}$), as shown in the top panels of Figure~\ref{fig3}. The VPD provides a useful visual diagnostic of the kinematic separation between cluster members and field stars. However, rather than relying on visual selection, we determine cluster membership and mean proper motion using probabilistic methods, specifically the Gaussian Mixture Model (GMM) described in the following subsection. 

We adopted two independent unsupervised clustering approaches to assess the robustness of membership determination. The motivation for employing both GMM and py\texttt{UPMASK} is twofold. First, different clustering algorithms rely on distinct statistical assumptions: GMM models the data as a mixture of multivariate Gaussian distributions in parameter space, while py\texttt{UPMASK} incorporates spatial clustering constraints via random-field rejection tests. Second, comparing independent probabilistic frameworks enables us to assess systematic differences in membership selection, particularly in regions with significant field-star contamination. Such cross-validation provides an internal consistency check before adopting a final membership sample for structural and dynamical analyses.

The bottom panels of Figure~\ref{fig3} show the corresponding color--magnitude diagrams (CMDs; $G$ vs. $G_{BP}-G_{RP}$), which further illustrate the separation between cluster members and field stars. The left panels display all stars within the selected cluster regions, the middle panels show high-probability cluster members identified using probabilistic membership analysis, and the right panels show field stars excluded from the membership sample. The combination of proper motion, parallax, and photometric information within a probabilistic framework enables reliable identification of genuine cluster members while minimizing contamination from field stars.

\begin{figure}[h!]
  \begin{minipage}[b]{0.42\textwidth}
    \includegraphics[width=\textwidth]{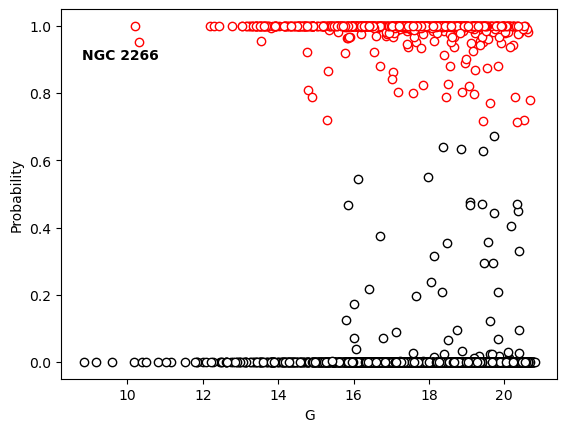}
  \end{minipage}
  \hfill
  \begin{minipage}[b]{0.42\textwidth}
    \includegraphics[width=\textwidth]{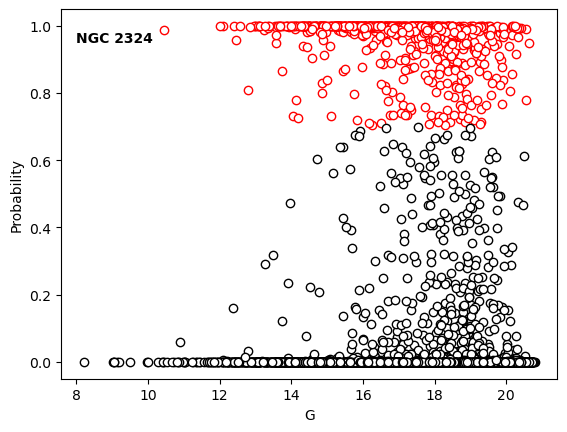}
  \end{minipage}
   \caption{Membership Probability as a function of $G$ magnitude, where the green circles represent member stars with membership probabilities greater than $70\%$ obtained using the GMM method, and the red circles represent member stars with membership probabilities greater than $70\%$ obtained using \texttt{pyUPMASK}. The orange circles indicate stars with more than $70\%$ common membership probabilities, while the black circles represent stars with less than $70\%$ common membership probabilities.}
   \label{fig5}
  \end{figure}

Identifying member stars is a crucial first step in analyzing OCs. Located along the Galactic plane, OCs often suffer from significant contamination by foreground and background field stars. This contamination can hinder the accurate determination of fundamental cluster parameters, including the PDMF. Therefore, strong and precise methods are required to distinguish true cluster members from field stars. Typically, member stars exhibit a higher spatial concentration toward the cluster center, forming a denser structure, while field stars are more randomly and uniformly distributed across the field, as illustrated in Figure \ref{fig5}.
To address this challenge, a VPD visually displays how cluster stars differ from field stars, as shown in Figure \ref{fig3}. The VPD and parallax distributions were used only as diagnostic tools to visualize the separation between cluster and field populations. Final membership probabilities were determined exclusively using the GMM algorithm. No hard VPD-radius or parallax cuts were applied in the final membership assignment. Moreover, mass is a key factor distinguishing member stars from field stars, as stars form from the collapse of gas and dust and ultimately reach the main sequence. In addition to the VPD, the Color-Magnitude Diagram (CMD) is vital for studying OCs, as it provides insights into the main sequence, turn-off stars, and blue stragglers that deviate from it. Building on these observational tools, clustering analysis can be employed to assign components to clusters such that components within a cluster are more similar to one another than those in other clusters \citep{bib95}. Consequently, we employed two unsupervised soft-clustering methods, the Gaussian Mixture Model (GMM) and \texttt{pyUPMASK}, to estimate membership probabilities for the OCs NGC 2266 and NGC 2324.\\

\begin{figure*}[h!]
\centering
    \includegraphics[height=7cm,width=7cm]{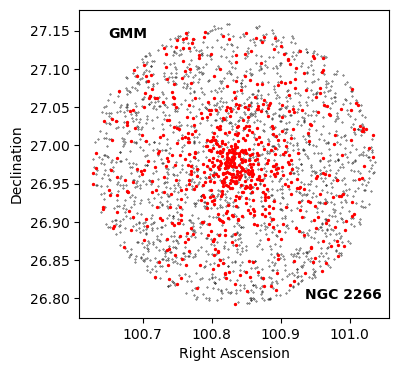}
  \hfill
    \includegraphics[height=7cm,width=7cm]{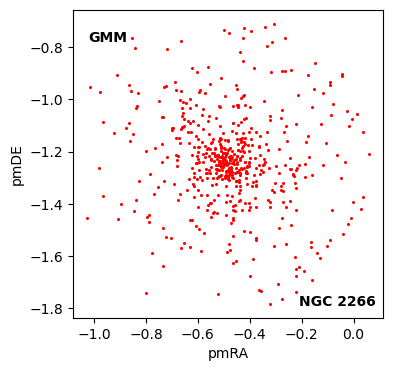}
  \hfill
    \includegraphics[height=7cm,width=4cm]{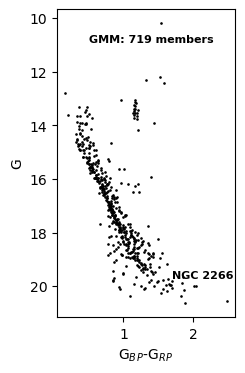}
    \includegraphics[height=7cm,width=7cm]{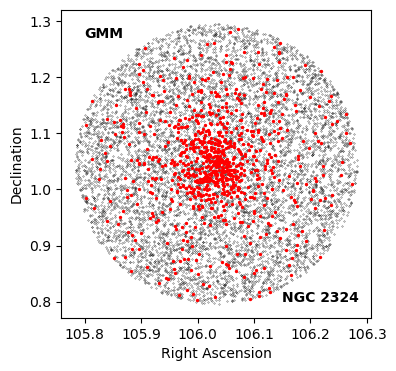}
  \hfill
    \includegraphics[height=7cm,width=7cm]{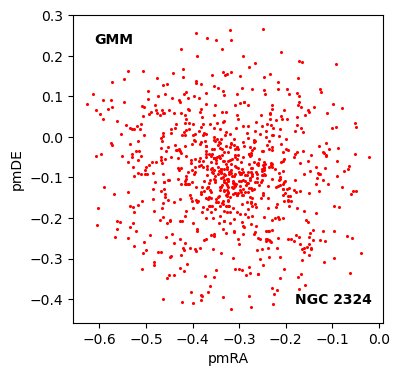}
  \hfill
    \includegraphics[height=7.09cm,width=4cm]{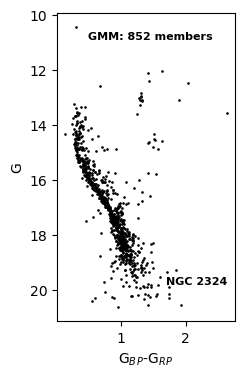}
  \caption{Cluster member identification using the Gaussian Mixture Model (GMM) method for NGC~2266 (top panels) and NGC~2324 (bottom panels). The left panels show the spatial distributions (RA vs.~DEC) of stars within the selected cluster regions, where red points indicate members with membership probability $P \geq 0.7$. The middle panels display the proper-motion diagrams ($\mu_{\alpha}\cos\delta$ vs.~$\mu_{\delta}$) for the identified members, illustrating their kinematic clustering. The right panels present the Gaia color--magnitude diagrams (CMDs) ($G$ vs.~$G_{\rm BP}-G_{\rm RP}$) for the high-probability members. A total of 719 and 852 cluster members are identified for NGC~2266 and NGC~2324, respectively.}
    \label{fig7}
\end{figure*}

\textbf{GMM} models the data as a mixture of multivariate Gaussian distributions. Unlike hard clustering methods, which assign each data point to a single cluster, GMM uses soft clustering. In soft clustering, each data point is assigned a probability of belonging to every cluster, rather than a binary assignment. This means each data point partially belongs to multiple clusters, with probabilities indicating the model's confidence in each group. This approach effectively captures data uncertainty and allows overlapping clusters. GMM is one of the most widely used clustering methods and is employed across various fields, including astrophysics \citep{bib91, bib90, bib93, bib94, bib92}. For example, it has been used to investigate the membership of two OCs, NGC 2244 \citep{bib96} and M 67 \citep{bib97}, based solely on proper-motion data. The method models data points as arising from a finite mixture of Gaussian distributions with unknown parameters, as noted by \citep{bib22}. These parameters are estimated using the expectation-maximization (EM) algorithm \citep{bib53}, an iterative method for maximizing the likelihood of probabilistic models. 
Firstly, we selected only five normalized astrometric parameters—RA, DEC, Parallax, PMRA, and PMDEC—from the dataset. Data normalization using these astrometric parameters is important because clustering outcomes can be affected by differing units and scales \citep{bib21,bib22}. After normalization, we applied the GMM clustering algorithm. In this step, we specifically used the \texttt{predict\_proba()} method to compute cluster membership probabilities. To further refine membership identification, we processed nearly 600 iterations. Following the iterations, we filtered stars using a 70\% probability cutoff. To assess the sensitivity of our results to the adopted probability threshold, we tested several cutoff values (60\%, 70\%, 80\%, and 90\%). Lower thresholds ($\le$ 60\%) significantly increased field-star contamination, particularly at faint magnitudes, broadening the main sequence and artificially flattening mass-function slopes. Conversely, higher thresholds ($\ge$80\%) reduced contamination further but at the expense of incompleteness in the low-mass regime, thereby biasing estimates of structural and dynamical parameters. The 70\% threshold provides an optimal compromise between completeness and purity, preserving the cluster sequence while maintaining adequate sampling across the adopted magnitude range. We identified 719 stars with probability $\geq$ 70 out of 3461 for NGC 2266, and 852 out of 11857 for NGC 2324, as shown in Figure \ref{fig7}.\\

{\bf {py\texttt{UPMASK}}} is a user-friendly, unsupervised Python-based clustering method that uses an astropy implementation to estimate membership probabilities. The method works well for stellar clusters with significant background contamination. \texttt{UPMASK} was first introduced in the R language \citep{bib46}. Later, a Python version called \texttt{\texttt{pyUPMASK}} \citep{bib48} appeared after the release of \textit{Gaia} EDR3 \citep{bib47}. It uses Ripley's K function algorithm \citep{bib49,bib50} to assess the proximity of the cluster to a uniform random distribution, including the required edge corrections for points near the domain boundaries. Hence, it checks whether a set of stars is uniformly distributed in space, and the RFR (Random Field Rejection) method is used to reject stars that are randomly distributed. The dataset is often contaminated by background or field stars that cannot be removed when they overlap with actual cluster members. Therefore, these stars are accepted. To clean this, the GUMM (Gaussian-uniform mixture model) is applied to a two-dimensional coordinate space, which is a simplified version of the spatial and proper-motion space modeling described in \citep{bib52}. There are no restrictions on the position, shape, or extent of the 2D Gaussian representing the cluster. The GMM, an iterative expectation-maximization (EM) algorithm (\citep{bib53}), is also used. It assigns each star a probability of belonging to a 2D Gaussian. The user can manually set a likelihood cut to select members and reject field stars. The method uses eight parameters: RA, DEC, parallax, and proper motions in RA and DEC, along with their uncertainties.

\begin{figure*}[h!]
    \includegraphics[height=7cm,width=7cm]{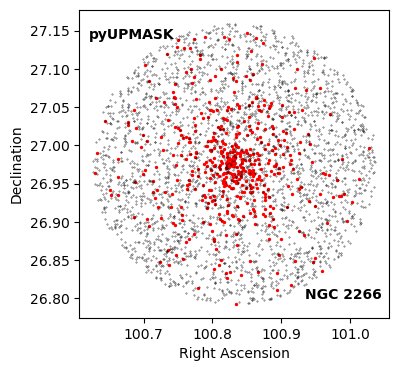}
  \hfill
    \includegraphics[height=7cm,width=7cm]{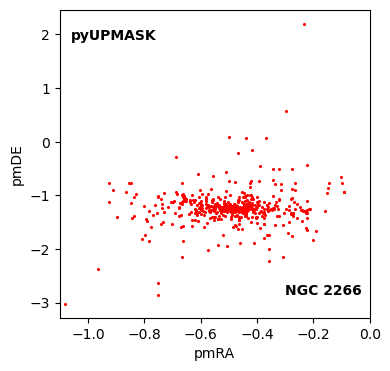}
  \hfill
    \includegraphics[height=7cm,width=4cm]{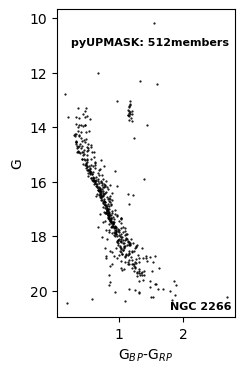}
   
\centering
    \includegraphics[height=7cm,width=7cm]{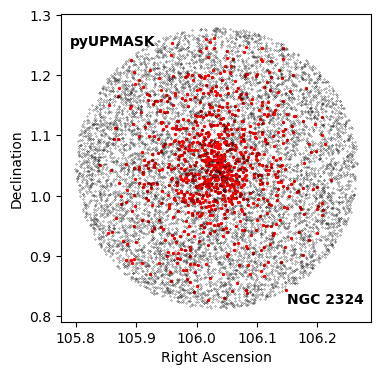}
  \hfill
    \includegraphics[height=7cm,width=7cm]{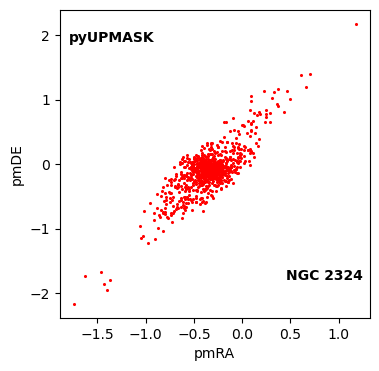}
  \hfill
    \includegraphics[height=7cm,width=4cm]{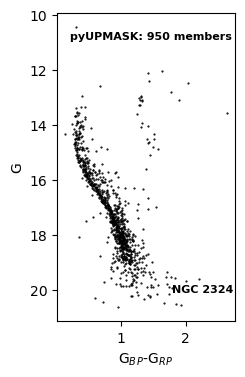}
  \caption{Same as Fig. 5 but using the \texttt{pyUPMASK} method.}
    \label{fig8}
\end{figure*}

Using the GMM method, we identified 719 and 852 members for NGC 2266 and NGC 2324, respectively, with membership probabilities exceeding 70\%, as shown in Figures \ref{fig7} and \ref{fig8}. In comparison, the \texttt{UPMASK} method obtained 512 members for NGC 2266 and 963 for NGC 2324, where the information containing magnitudes was only available for 950 stars. However, \texttt{UPMASK}'s membership data, as highlighted in the Color-Magnitude diagrams, exhibited significant background contamination from field stars. In contrast, diagrams based on GMM results demonstrated clearer separation. To quantitatively compare the two clustering methods, we first examine the total number of high-probability members ($P \geq 0.7$) identified by each technique. For NGC 2266, the GMM method identifies 719 members, whereas py\texttt{UPMASK} identifies 512. For NGC 2324, GMM identifies 852 members, whereas py\texttt{UPMASK} selects 963. The GMM sample retains a well-defined main sequence over the adopted magnitude range while maintaining lower CMD dispersion than pyUPMASK. In contrast, although py\texttt{UPMASK} identifies a comparable number of members in NGC 2324 (963 stars), the corresponding CMD (Figure 6) shows a visibly greater scatter around the main-sequence and turn-off regions. This increased dispersion indicates greater residual-field contamination, particularly at fainter magnitudes. The difference between the two methods is therefore not merely in total member counts, but in the photometric coherence and structural consistency of the selected samples.  To quantify the photometric coherence of the selected members, we restricted the analysis to stars in the visually identified main-sequence region of the Gaia CMD, with $14 \leq G \leq 19$ mag. Obvious evolved stars, blue straggler candidates, and outliers were excluded. The standard deviation of the $(G_{\rm BP}-G_{\rm RP})$ color distribution was then computed for these selected main-sequence stars. For NGC~2266, the standard deviation is 0.23 for the GMM-selected sample and 0.30 for the pyUPMASK sample. Similarly, for NGC~2324, the dispersion is 0.24 for GMM and 0.40 for pyUPMASK. These results indicate that the GMM-selected members define a narrower and more coherent main sequence compared to pyUPMASK, implying lower residual field-star contamination. In addition to the CMD dispersion analysis, we compared our membership results with the Gaia DR3 catalog of \citet{hunt2023improving}. The stars identified in common between the two studies are predominantly located along the well-defined main-sequence and turn-off regions of the CMD, indicating broad consistency in identifying the principal cluster population. Differences between the membership lists are expected due to variations in adopted magnitude limits, search radii, membership selection criteria, and underlying clustering methodologies. In particular, our analysis uses a deeper Gaia DR3 sample ($G \leq 20$ mag) and a probabilistic Gaussian Mixture Model (GMM) framework that incorporates astrometric information into a multidimensional parameter space, whereas \citet{hunt2023improving} employed a density-based clustering approach.

Rather than serving as a direct validation of individual memberships, the comparison provides an independent consistency check of the overall cluster population. The common members recovered by both studies occupy the expected cluster sequences in the CMD and exhibit coherent astrometric properties, providing an additional consistency check of the adopted membership selection. Furthermore, the GMM-based sample produces a tighter main sequence and lower CMD dispersion than the pyUPMASK sample, indicating reduced residual field-star contamination. Consequently, we adopt the GMM membership catalogue, comprising 719 and 852 high-probability members for NGC~2266 and NGC~2324, respectively, for all subsequent structural, photometric, and dynamical analyses.

\subsection{Comparison of Our Membership Results with Hunt \& Reffert et al. (2024) and Cantat-Gaudin et al. (2020):} 
\begin{figure*}[h!] 
  \begin{minipage}[b]{0.51\textwidth}
    \includegraphics[width=\textwidth]{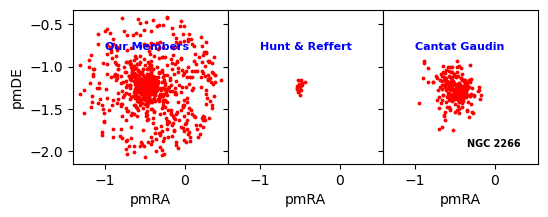}
  \end{minipage}
  \hfill
  \begin{minipage}[b]{0.5\textwidth}
    \includegraphics[width=\textwidth]{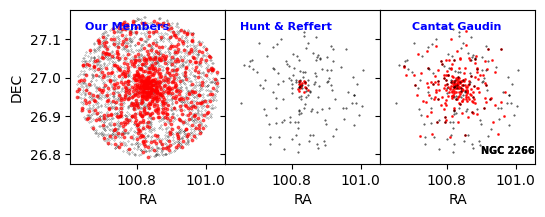}
  \end{minipage}
  \begin{minipage}[b]{0.51\textwidth}
    \includegraphics[width=\textwidth]{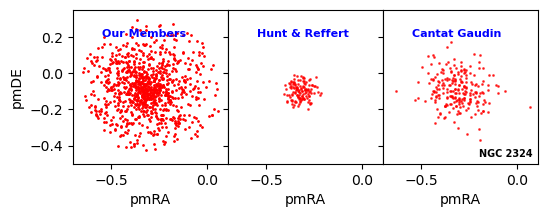}
  \end{minipage}
  \hfill
  \begin{minipage}[b]{0.5\textwidth}
    \includegraphics[width=\textwidth]{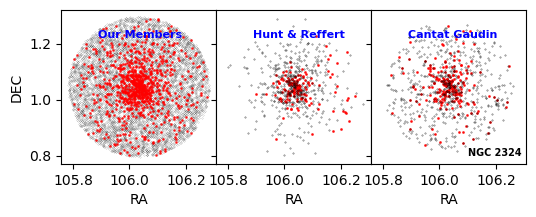}
  \end{minipage}
  \caption{Comparison of the proper-motion and spatial distributions of cluster members for NGC~2266 (top panels) and NGC~2324 (bottom panels). The left panels show the proper-motion diagrams ($\mu_{\alpha}\cos\delta$ vs. $\mu_{\delta}$) for member stars identified in this work using the GMM method with membership probability $P \geq 0.7$. The middle panels present the members reported by Hunt \& Reffert (2023), while the right panels show members listed by Cantat-Gaudin et al.~(2020). The corresponding spatial distributions (RA vs.~DEC) are also shown for comparison. A total of 719 and 852 high-probability members ($P \geq 0.7$) are identified for NGC 2266 and NGC 2324, respectively.}
    \label{fig:pmra}
\end{figure*}

\begin{figure*}[h!]
\vspace{0.5cm}
\hspace{0.9cm}
  \begin{minipage}[b]{0.4\textwidth}
    \includegraphics[width=1.11\textwidth]{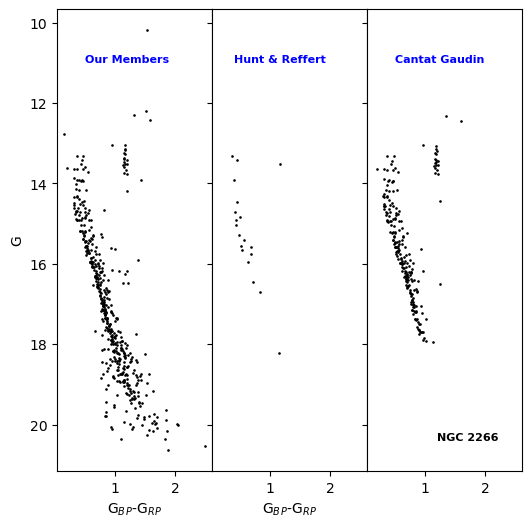}
  \end{minipage}
  \hspace{1.2cm}
  \begin{minipage}[b]{0.4\textwidth}
    \includegraphics[width=1.11\textwidth]{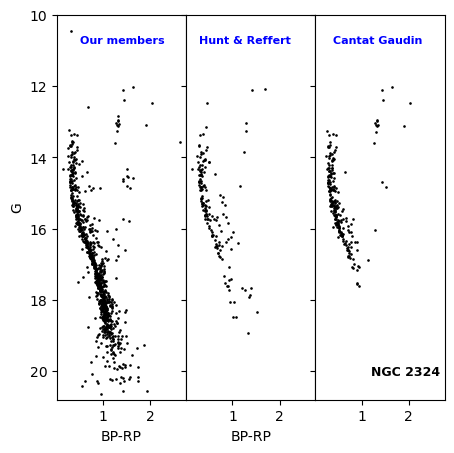}
  \end{minipage}
\caption{Comparison of the Gaia color--magnitude diagrams (CMDs) for the OCs NGC~2266 (left) and NGC~2324 (right). The panels show the CMDs ($G$ vs.~$G_{\rm BP}-G_{\rm RP}$) for cluster members identified in this work using the GMM method (left panels), members reported by Hunt \& Reffert (2023) (middle panels), and members from Cantat-Gaudin et al.~(2020) (right panels). The deeper Gaia DR3 sample and adopted membership methodology result in a larger number of probable low-luminosity members compared to previous catalogs.}
    \label{fig:10}
\end{figure*}

Using \textit{Gaia}~DR3 data and GMM-based membership analysis, we identified 719 members for NGC~2266 and 852 for NGC~2324. In comparison, Cantat-Gaudin (2020) reported only 227 members in NGC~2266 and 225 members in NGC~2324 with membership probabilities above 70\%, based on \textit{Gaia}~DR2 and the same search radius. We also compared our membership results with the Gaia DR3 catalog of Hunt \& Reffert (2023), which employed a density-based clustering algorithm (DBSCAN). Their catalogs report approximately 332 members for NGC 2266 and 632 for NGC 2324 within comparable search radii. The larger number of members identified in our analysis is primarily due to the deeper photometric selection (G $\le$ 20 mag) and the probabilistic treatment of astrometric parameters. The comparison demonstrates broad consistency between the different membership catalogs despite methodological differences and should be interpreted primarily as an external consistency check rather than a direct validation of the adopted membership sample. The distribution of stars in the proper-motion plane provides a robust diagnostic for membership determination. Previous studies have successfully applied the GMM method using only proper-motion data to determine memberships in open clusters \citep{bib96,bib97}. Similarly, the proper-motion distributions of our clusters (Figure~\ref{fig:pmra}) show a well-defined concentration, supporting the robustness of the adopted membership selection. To precisely determine the Mean Proper motion of both clusters, we selected only members with membership probabilities greater than 70\%. By fitting the 1-D Gaussian function on each of their histograms, we obtain values for mean proper motions for both clusters in the directions of RA and DEC as: ( -0.488 ± 0.021 mas/yr \& -1.247 ± 0.024 mas/yr) for NGC 2266 and (- 0.325 ± 0.005 mas/yr \& - 0.089 ± 0.008 mas/yr) for NGC 2324. These mean values of proper motions agree with \citep{bib43} and \citep{bib44}. These stars have PM errors of $\leq$ 0.021 for NGC 2266 and $\leq$  0.008 for NGC 2324. The CMD of a stellar cluster is an effective tool for estimating its age, distance, and reddening. It is also helpful in distinguishing member stars from field stars along the cluster's main sequence. For stars with membership probabilities greater than 70\%, the high-probability members identified by \citet{bib20} are largely recovered within our Gaussian Mixture Model (GMM) membership sample, as shown in Figure~\ref{fig:10}. The common members occupy the expected cluster sequences in the CMD and exhibit similar astrometric properties, demonstrating broad consistency between the two membership determinations despite differences in the adopted datasets and methodologies. Our study is two magnitudes deeper than Cantat-Gaudin, making it advantageous for analyzing the cluster mass function.

\section{Parallax-Based Distance Estimation }\label{sec5}

\begin{figure*}[h!]
  \begin{minipage}[b]{0.48\textwidth}
    \includegraphics[width=\textwidth]{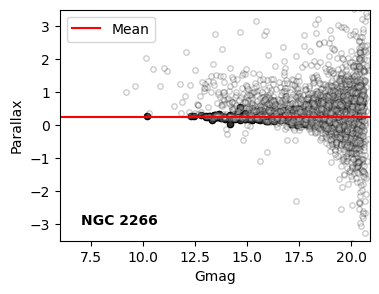}
  \end{minipage}
  \hfill
  \begin{minipage}[b]{0.5\textwidth}
    \includegraphics[width=\textwidth]{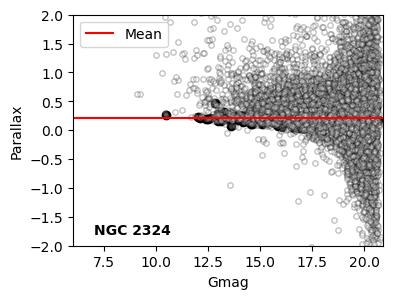}
  \end{minipage}
\caption{Parallax as a function of Gaia $G$ magnitude for probable member stars of NGC~2266 (left panel) and NGC~2324 (right panel). The grey circles represent all stars in the selected cluster regions, while the black points indicate high-probability cluster members ($P \geq 0.7$) identified using the GMM method. The horizontal red line marks the mean parallax value derived from the Gaussian fit to the parallax distribution of the member stars. This visualization illustrates the concentration of cluster members around the mean parallax and highlights the increasing parallax uncertainties toward fainter magnitudes.}
    \label{plx}
\end{figure*}

Cluster distances were estimated using the trigonometric parallaxes of high-probability member stars. The distribution of parallaxes as a function of Gaia G magnitude for the high-probability cluster members is shown in Figure~\ref{plx}, illustrating the concentration of members around the mean cluster parallax and the increase in parallax uncertainties toward fainter magnitudes. We first selected stars with membership probabilities exceeding 70\% from the Gaussian Mixture Model (GMM) analysis. A Gaussian profile was fitted to the parallax distributions of these members to determine the mean cluster parallaxes. The resulting mean parallaxes are $0.253\pm0.010$ mas for NGC 2266 and $0.210\pm0.008$ mas for NGC 2324. These values are consistent with the mean parallaxes reported in previous catalogs \citep{bib44,bib43}, also mentioned in the Table. \ref{table5}.

Gaia DR3 parallaxes include a known systematic offset; therefore, we applied the global parallax zero-point correction of $-$0.029 mas following Lindegren et al. (2021). While the distance to a star can formally be estimated by directly inverting the parallax ($d = 1/\varpi$), this method becomes unreliable when the relative parallax uncertainty is significant, particularly for distant clusters with small parallax values. In such cases, direct inversion can introduce biases and unrealistic distance estimates.

To obtain more reliable distances, we used the probabilistic distance estimates from \citet{bailer2018estimating}. This method uses a Bayesian framework that combines the measured trigonometric parallax with its associated uncertainty and assumes an exponentially decreasing space-density prior. Such an approach has been shown to provide more robust distance estimates when parallax uncertainties are non-negligible.

Using the mean parallaxes obtained in this work, the direct inversion method yields trigonometric distances of 3.95 kpc for NGC 2266 and 4.76 kpc for NGC 2324. However, the corresponding Bayesian distances from the Bailer-Jones catalog are $3.55\pm0.23$ kpc for NGC 2266 and $4.18\pm0.24$ kpc for NGC 2324, which we adopt for all subsequent analyses. These values are consistent within uncertainties with previously reported distances in the literature (e.g., \citet{cantat2020painting} and \citet{hunt2023improving}).

\section{Radial Density and Structural Parameters of Clusters} \label{rdp} 
\begin{figure*}[h!]
  \begin{minipage}[b]{0.41\textwidth}
    \includegraphics[width=\textwidth]{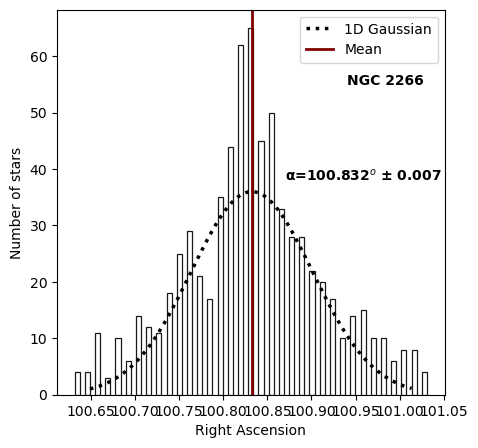}
  \end{minipage}
  \begin{minipage}[b]{0.4\textwidth}
    \includegraphics[width=\textwidth]{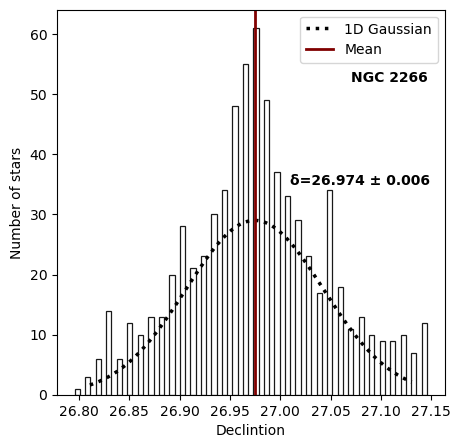}
  \end{minipage}
  \begin{minipage}[b]{0.4\textwidth}
    \includegraphics[width=\textwidth]{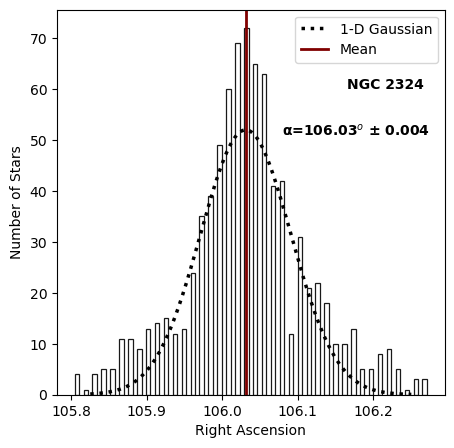}
  \end{minipage}
  \hfill
  \begin{minipage}[b]{0.41\textwidth}
    \includegraphics[width=\textwidth]{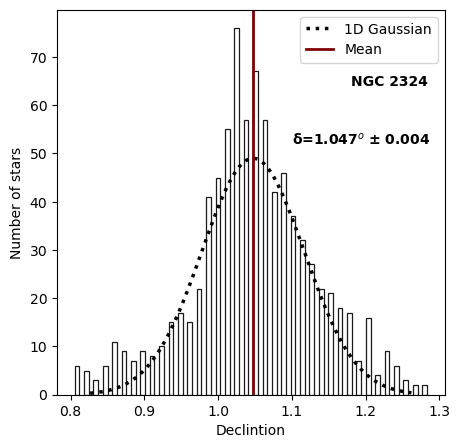}
  \end{minipage}
\caption{Histograms of stellar number density along right ascension (left panels) and declination (right panels) for the clusters NGC~2266 (top row) and NGC~2324 (bottom row). The distributions were constructed using probable cluster members within the adopted cluster region. The black dashed curves represent the best-fitting one-dimensional Gaussian profiles to the stellar count distributions. The vertical red lines indicate the mean positions derived from the Gaussian fits, which correspond to the estimated cluster centers. The derived central coordinates are $\alpha = 100.832^{\circ} \pm 0.007^{\circ}$ and $\delta = 26.974^{\circ} \pm 0.006^{\circ}$ for NGC~2266, and $\alpha = 106.03^{\circ} \pm 0.004^{\circ}$ and $\delta = 1.047^{\circ} \pm 0.004^{\circ}$ for NGC~2324.}
    \label{fig12}
\end{figure*}

The spatial structure of an open cluster is crucial for understanding the distribution of stellar masses among its members. To construct the radial density profile of a cluster, it is necessary first to determine its center. In earlier studies of OCs, the cluster center was determined by visual inspection \citep{bib225,bib224}. In our study, the cluster center was estimated using the star-count method, applied to stars with membership probabilities $\geq 70\%$. One-dimensional Gaussian fits were applied to the histograms of the distributions of the right ascension (RA) and declination (DEC) of the probable member stars to determine the coordinates of the cluster center, as shown in Figure~\ref{fig12}.
 The cluster centers were determined to be the calculated mean values with uncertainties: (100.832 ± 0.007 deg, 26.974 ± 0.006 deg) for NGC 2266 and (106.031 ± 0.005 deg, 1.047 ± 0.005 deg) for NGC 2324. These values align well with previous studies \citep{bib44, bib43}. 

The empirical model of \citep{bib42} was also fitted over radial density profiles to estimate the structural parameters of the clusters. The King's model equation can be expressed as follows:\\

 ~~~~~~~~~~~~~~~~~~f(r)= f$_{bg}$ + f$_0$ ( $\frac{1}{\sqrt{1 + (r / r_c)^2}}$ - $\frac{1}{\sqrt{1 + (r / r_t)^2}}$ )$^2$ \\

\begin{figure*}[h!]
  \begin{minipage}[b]{0.49\textwidth}
    \includegraphics[width=\textwidth]{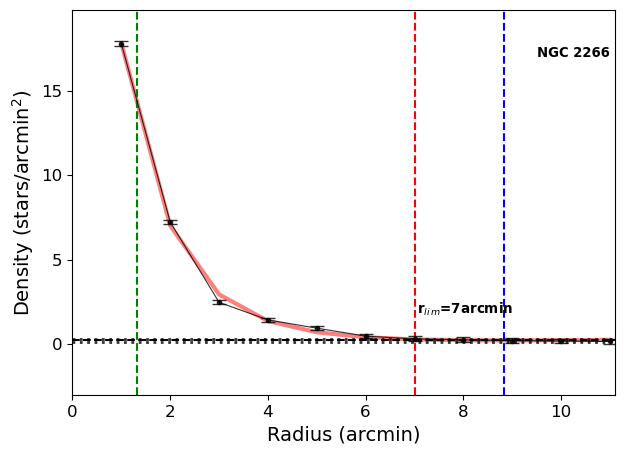}
  \end{minipage}
  \hfill
  \begin{minipage}[b]{0.49\textwidth}
    \includegraphics[width=\textwidth]{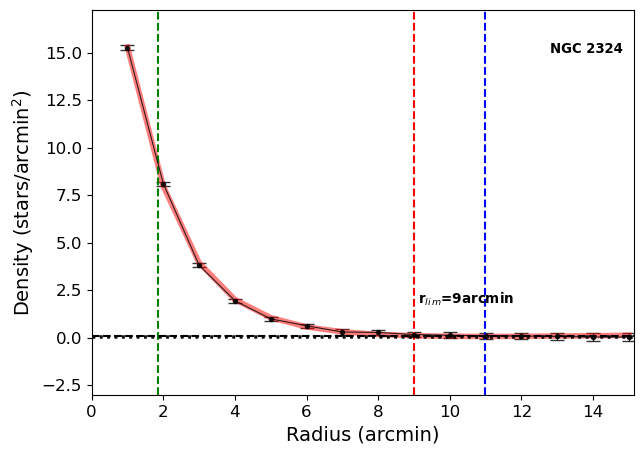}
  \end{minipage}
    \caption{Radial surface density profiles of the OCs NGC~2266 (left panel) and NGC~2324 (right panel). The stellar surface density (stars arcmin$^{-2}$) is shown as a function of radial distance from the cluster center. Black points represent the observed stellar densities in concentric annuli, with Poisson uncertainties derived from sampling statistics. The solid red curve indicates the best-fitting King profile to the observed density distribution. The horizontal dashed line marks the estimated background field density, while the dotted line represents the uncertainty in the background level. The green vertical dashed line denotes the core radius ($r_c$), the red vertical dashed line indicates the limiting radius ($r_{\mathrm{lim}}$), and the blue vertical dashed line marks the tidal radius ($r_t$) of the clusters.}
    \label{fig13}
\end{figure*}

Where f$_{bg}$, f$_0$, and r$_c$ are the clusters' background density, maximum stellar density, and core radius, while r$_t$ is the tidal radius. From here, the concentration parameter (c) = log($\frac{r_t}{r_c}$) is determined to be 0.82 for NGC 2266 and 0.77 for NGC 2324. The concentration parameter measures the degree of central concentration of the cluster. Higher values of c indicate a more centrally condensed structure, suggesting stronger gravitational binding and more advanced dynamical evolution. Clusters with higher concentration parameters are generally more dynamically relaxed and less susceptible to tidal disruption, while lower values indicate more diffuse stellar distributions. These structural parameters serve as essential inputs for subsequent dynamical analyses, including mass segregation and relaxation-time calculations. Furthermore, the limiting radius, representing the boundary where the cluster's stellar density merges with the background, was consistent with prior determinations. We obtained limiting radii of 7 arcmin for NGC 2266, aligning well with the values noted by \citep{bib56}, and nine arcmin for NGC 2324, in good agreement with the 8.9 arcmin noted by \citep{bib113}, but is almost double the value provided by \citep{bib220} for NGC 2266 and 1/3 greater for NGC 2324. In Figure (\ref{fig13}), the radial density plots, along with the King model fit, show a smooth decline in stellar density from the center outward, confirming the reliability of the structural parameter estimates. These parameters, as listed in Table \ref{table1}, serve as essential inputs for subsequent dynamical studies, including mass segregation and relaxation-time calculations, thereby providing insight into the clusters' evolutionary state.

\begin{table*}[ht]
\caption{ Structural parameters derived from the radial density profile analysis for the OCs NGC~2266 and NGC~2324. The table lists the adopted field size, central stellar density ($f_0$), background density ($f_{\rm bg}$), core radius ($r_c$), limiting radius ($r_{\rm lim}$), tidal radius ($r_t$), and concentration parameter ($c = \log(r_t/r_c)$) obtained from the King profile fitting.} 
\centering 
\vspace{0.15cm}
\begin{tabular}{c c c c c c c c} 
\hline \hline 
Cluster & Field size &  f$_0$ &f$_{bg}$ & r$_c$ & r$_{lim}$    & r$_t$ & c  \\ [0.1ex] 
Name &  (arcmin) & & & (pc) & (pc) & (pc)\\
\hline 

NGC 2266 & 11' $\times$ 11' & 41.70 $\pm$ 1.86 & 0.22 $\pm$ 0.09 & 1.36 $\pm$ 0.011 & 7.23 $\pm$ 0.47 & 9.13 $\pm$ 1.44 & 0.82 \\ 
NGC 2324 & 15' $\times$ 15'& 29.87 $\pm$ 0.64 & 0.07 $\pm$ 0.03 & 2.26 $\pm$ 0.05 & 10.94 $\pm$ 0.63 & 13.34 $\pm$ 0.68 & 0.77\\ [0.1ex] 
\hline 
\end{tabular}
\label{table1} 
\end{table*}

\section{Determination of Cluster Parameters} \label{7}

\subsection{Reddening} \label{redenn}

Interstellar reddening causes starlight to appear redder than its intrinsic color due to the absorption and scattering from interstellar dust. Accurate reddening estimates along a cluster's line of sight are essential, as they directly affect cluster age and distance measurements. The interstellar reddening for both clusters was determined by color excess \textit{E(G$_{BP}$-G$_{RP}$)} through isochrone fitting to the observed CMD. This method helps us to estimate the color shift caused by interstellar dust extinction. The derived color excess values range from 0.11 to 0.31. For the evaluation of extinction in the Gaia G band, we employed the empirical relation A$_G$ = 1.86 $\times$ \textit{E(G$_{BP}$-G$_{RP}$)}, as described in \citep{bib232,bib226}.
To quantify reddening towards NGC 2266 and NGC 2324, we used (J-H) versus (J-K) two-color diagrams (TCDs), created from 2MASS photometry for the most probable main-sequence cluster members. We visually fitted the intrinsic zero-age main sequence (ZAMS) of solar metallicity \citep{bib4} to the observed TCDs. The fitting was performed using a visual best-fit approach, shifting the intrinsic ZAMS along the reddening vector until it agreed optimally with the observed main-sequence stars. Only high-probability cluster members (P $\ge$ 0.7) were used to avoid field-star contamination. The best fit was selected through visual inspection, identifying the reddened ZAMS that best matched the observed main-sequence distribution. This approach is commonly used in open-cluster reddening studies and yields robust estimates of reddening when the main sequence is well-defined. In each TCD, the intrinsic ZAMS appears as a solid line; the reddened ZAMS, as a dotted line, is shifted along the reddening vector. For NGC 2266, we used 422 members, and for NGC 2324, 566 members, obtained by matching RA \& DEC objects from 2MASS \citep{bib100} with our probable members from Gaia DR3 \citep{bib8}, all with P $\geq$ 0.7. For $E(B-V)$, the ZAMS was shifted across a range of reddening values. The ratio of $E(J-H)$ to $E(J-K)$ values used to shift the ZAMS should be between 0.55 and 0.60, as mentioned in \citep{bib26}. The values added to shift the ZAMS are then put in the relations mentioned in \citep{bib31}:\\ 

\begin{center}
E($J-H$)= 0.309 $\times$ E(B-V)\\
E($J-K$)=  0.48 $\times$ E(B-V)\\
\end{center}

\begin{figure}[h!]
  \begin{minipage}[b]{0.42\textwidth}
    \includegraphics[width=\textwidth]{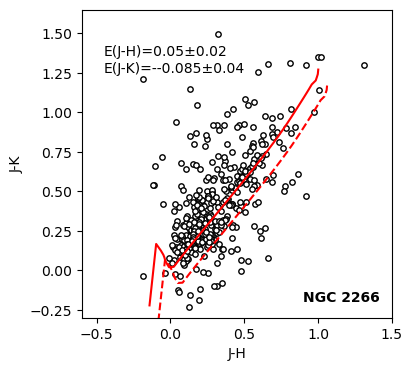}
  \end{minipage}
  \hfill
  \begin{minipage}[b]{0.42\textwidth}
    \includegraphics[width=\textwidth]{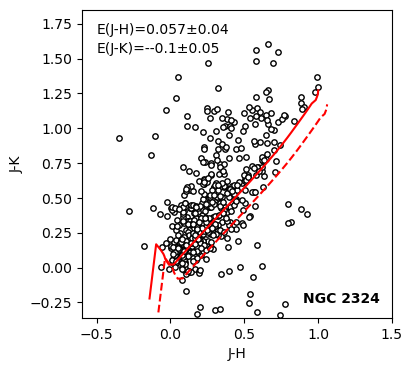}
  \end{minipage}
\caption{Near-infrared $(J-H)$ versus $(J-K)$ color–color diagrams for the OCs NGC~2266 (top panel) and NGC~2324 (bottom panel) using probable cluster members. The black circles represent the observed stellar colors from 2MASS photometry. The solid red curve denotes the intrinsic Zero-Age Main Sequence (ZAMS) adopted from Caldwell et al. (1993). The dashed red curve represents the same ZAMS shifted along the reddening vector to match the observed stellar distribution. The displacement between the intrinsic and shifted sequences yields the color excess values, $E(J-H)$ and $E(J-K)$, for each cluster.}
    \label{fig14}
\end{figure}

The shifts in $(J-H)$ and $(J-K)$ followed the standard interstellar reddening law, which constrains the ratio $E(J-H)/E(J-K)$ to lie within the range 0.55--0.60. We measured ratios of 0.58 and 0.57 for NGC~2266 and NGC~2324, respectively, in excellent agreement with the expected values. Using the shifted ZAMS and the empirical relations of \citet{bib220}, we converted the near-infrared colour excesses into optical reddening estimates. This procedure yielded $E(B-V)=0.17 \pm 0.04$ mag for NGC~2266, consistent with the values reported by \citet{bib55,bib56}, and $E(B-V)=0.22 \pm 0.06$ mag for NGC~2324, in close agreement with the value of 0.23 mag reported by \citet{bib221}, as shown in Figure~\ref{fig14}. Although some differences remain compared with other estimates in the literature (Table~\ref{table5}), the derived reddening values are generally consistent within their uncertainties. The quoted uncertainties reflect the combined effects of photometric errors and the subjective uncertainty associated with the visual ZAMS-fitting procedure. The use of high-probability cluster members ($P \geq 0.7$) minimizes field-star contamination and improves the reliability of the reddening determination. The reddening estimates obtained from the 2MASS TCD/ZAMS fitting serve as an independent consistency check. For the subsequent isochrone fitting and derivation of fundamental cluster parameters, we adopt the reddening values derived from the Gaia CMD analysis, which are also used in the luminosity-function and mass-function analyses.

\subsection{Metallicity} 

The metallicity of open clusters provides important insights into their formation and chemical evolution. To compare the adopted photometric metallicities with available spectroscopic measurements, the high-probability GMM members ($P \geq 0.7$) were cross-matched with the LAMOST DR7 catalogue using their equatorial coordinates (RA and Dec). A total of 53 high-probability members of NGC~2266 were successfully cross-matched with the LAMOST DR7 catalogue. Of these, 48 stars had reported [Fe/H] measurements and were included in the spectroscopic metallicity analysis, while the remaining 5 matched stars lacked metallicity estimates and were therefore excluded. No additional quality filtering (e.g., signal-to-noise or quality-flag selection) or outlier rejection was applied beyond the requirement that the stars be high-probability cluster members with available LAMOST [Fe/H] measurements. The spectroscopic metallicities were converted to the corresponding heavy-element abundance ($Z$) following \citet{bib41}:

\begin{center}
Z = $\frac{0.013}{(0.04) \times 10^{-(Fe/H)}}$ 
\end{center}

Although the metallicity derived from LAMOST spectroscopy provides an independent estimate, the adopted metallicity for the isochrone fitting was chosen to yield the best photometric fit to the Gaia CMD.

The metallicity of each cluster was determined through isochrone fitting to the Gaia DR3 colour--magnitude diagram (CMD). For NGC~2266, previous studies reported metallicities of $Z=0.004$ \citep{bib55,bib56} and $Z=0.007$ \citep{bib57,bib54}. Similarly, metallicity estimates for NGC~2324 range from $Z=0.004$ \citep{bib2} to $Z=0.019$ \citep{bib60}, with intermediate values of $Z=0.012$ reported by \citet{bib59,bib54}. We examined these literature values and found that they do not adequately reproduce the observed morphology of the Gaia DR3 CMD, particularly in the main-sequence and turnoff regions.

The best agreement between the observed CMDs and the theoretical isochrones was obtained for $Z=0.0084$ ([Fe/H]$=-0.35$ dex) in NGC~2266 and $Z=0.0038$ ([Fe/H]$=-0.70$ dex) in NGC~2324. These metallicities were therefore adopted to determine the cluster parameters and used in all subsequent analyses.

As an independent spectroscopic consistency check, the high-probability cluster members ($P \geq 0.7$) were cross-matched with the LAMOST DR7 catalogue \citep{bib98}. For NGC~2266, 53 high-probability members were successfully matched with the LAMOST DR7 catalogue; however, [Fe/H] measurements were available for only 48 stars. These 48 stars yield a mean metallicity of [Fe/H]$=-0.62$ dex with a standard deviation of 0.57 dex. The relatively large dispersion likely reflects the heterogeneous quality of the available spectroscopic measurements, the intrinsic uncertainties in individual stellar metallicities, and possible residual contamination.

\subsection{Age and Distance Estimation Through Isochrone Fitting} \label{sec7.3}

\begin{figure*}[h!]
  \begin{minipage}[b]{0.45\textwidth}
    \includegraphics[width=\textwidth]{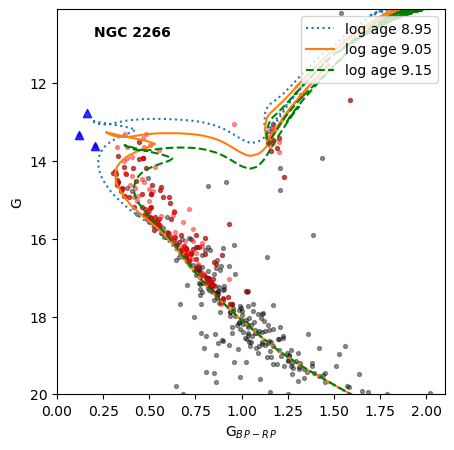}
  \end{minipage}
  \hfill
  \begin{minipage}[b]{0.45\textwidth}
    \includegraphics[width=\textwidth]{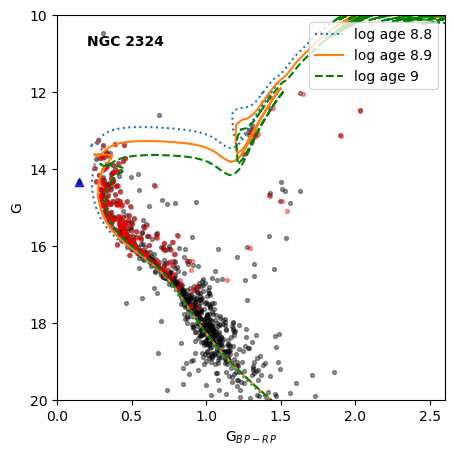}
  \end{minipage}
\caption{All the stars here are probable members with a probability above 70 percent. The curves are isochrones from \citep{bib61}, with log ages 8.95, 9.05, and 9.15 for NGC 2266, and 8.8, 8.9, and 9.0 for NGC 2324. The red stars denote 194 members for NGC 2266 and 183 members for NGC 2324, matched with \cite{bib43} members with above 70 percent membership probability. The grey solid dots denote members newly obtained from \citep{bib8} membership under a 70 percent cutoff. The blue triangles represent blue straggler stars in NGC 2266. A small fraction of stars is scattered toward the top-right and bottom-left regions of the CMD are likely residual field-star contamination, unresolved binaries, or stars with larger photometric uncertainties that remain in the probabilistic membership sample.}
    \label{fig15}
\end{figure*}

Accurate estimates of distance and age are essential for investigating Galactic structure and chemical evolution using OCs (OCs). To achieve this, color–magnitude diagrams (CMDs) were created using \textit{Gaia} $G$-band magnitudes and $(G_{BP}-G_{RP})$ colors for the most probable cluster members. For NGC 2266, 719 evolved member stars with available photometry were used, while 852 members were analyzed for NGC 2324. These CMDs clearly show the main sequence, the turn-off point, and a distinct group of red giant stars in NGC 2266, as observed in previous studies. In addition to \citep{bib43}, we have deeper coverage, resulting in improved member selection and a better chance of fitting isochrones. Isochrone fitting used the models of \citet{bib61}. For NGC 2266, the best-fitting isochrones span log(age) 8.95–9.15, with a mean of 9.05. For NGC 2324, the best-fit range is log(age) 8.8–9.0, with a mean of 8.9. The isochrone with log(age) = 9.05 and $Z = 0.0084$ best matches NGC 2266, while log(age) = 8.90 and $Z = 0.0038$ fit NGC 2324, as shown in Figure~\ref{fig15}. Mean $G$-band extinction values, $A_G$, are 0.17 for NGC 2266 and 0.22 for NGC 2324, considering only members with probabilities over 70\%.   From the CMDs ($(G_{BP}-G_{RP})$ vs. $G$) based on \citet{bib8}, we derived the distance moduli and ages of both clusters. The distance moduli yield heliocentric distances of 3.16 kpc for NGC 2266 and 3.98 kpc for NGC 2324, matching previous studies \citep{bib114,bib55,bib221,bib220,bib54,bib209}. The discrepancy between the isochrone-derived and parallax-based distances may arise from the well-known degeneracy among reddening, metallicity, age, and distance during isochrone fitting. Small variations in these parameters can produce comparable CMD morphologies and, therefore, slightly different distance estimates. Best-fit isochrones give ages of $1.1 \pm 0.1$ Gyr for NGC 2266 and $790 \pm 150$ Myr for NGC 2324, consistent with earlier results \citep{bib205,bib55,bib221,bib56,bib54,bib209}. Although the derived ages and distances are broadly consistent with previous studies, the present analysis benefits from Gaia DR3’s deeper photometric coverage and improved membership determination using the GMM method. This allows us to trace the cluster main sequence to significantly fainter magnitudes than in earlier works, resulting in a better-defined CMD and a more reliable identification of low-mass members. The improved faint-end coverage provides stronger constraints on isochrone fitting and enables more accurate determination of cluster parameters, particularly for subsequent luminosity- and mass-function analyses. We derived the Galactocentric coordinates using the adopted Bailer--Jones Bayesian distances (Section~\ref{sec5}) and the Galactic longitudes and latitudes of the clusters. Assuming the Sun is 8.3 kpc from the Galactic center \citep{bib227}, the coordinates are (-3440.9 $\pm$ 224, -471.3 $\pm$ 31, 631.1 $\pm$ 41) pc for NGC 2266 and (-3490.3 $\pm$ 200, -2305.5 $\pm$ 132, 240.9 $\pm$ 14) pc for NGC 2324, agreeing with \citet{bib43}. The Galactocentric radii ($R_{GC}$), using \citet{bib217}, are 11.95 kpc for NGC 2266 and 12.21 kpc for NGC 2324, also matching \citet{bib43}.

\subsection{Blue Straggler Stars (BSS):}  
 
BSS are stars that appear more luminous and bluer than other stars \citep{bib234}. These stars are hot, blue, and massive, and follow different evolutionary paths from the turn-off point. BSS are identified by their position in a CMD; if stars extend the main sequence above the turnoff in brightness, they are blue stragglers. They usually appear at brighter magnitudes and hotter temperatures than other stars in the CMD, forming an extension of the main sequence \citep{bib4,bib234}, distributed between the turnoff and up to two magnitudes brighter. In some clusters, the brightest are up to 3 magnitudes above the turnoff. The lower-luminosity limit of each OC is not always clear; sometimes, stars appear bluer and fainter than the turnoff on the zero-age main sequence, making them seem younger and classifying them as blue stragglers.\\

At first, BSS were detected and verified by  \citep{bib101} in the CMD of globular cluster M3. This detection appeared as a broadening of the cluster's main sequence, extending blueward and above the main-sequence turnoff. Generally, BSS are found in globular clusters in the Milky Way Galaxy \citep{bib102,bib103} or in dwarf spheroidal galaxies \citep{bib104}. They are also present in intermediate-aged to old-aged OCs \citep{bib105,bib106}, as mentioned in \citep{bib107}.
OCs consist of groups of stars that originate from the same molecular cloud. These stars share the same chemical composition, kinematics, and dynamics. This common origin is a clear cause of BSS. Sometimes, mass transfer from a binary companion can regenerate the acceptors. This process can lead to the emergence of BSS \citep{bib108}. Discrete stellar mergers, resulting from definite stellar collisions, are also related to the origin of BSS \citep{bib109,bib110,bib111}. Another study assessed how magnetic winds remove angular momentum from the main-sequence of tidally coincident binaries. It is recommended that this process accounts for at least $\frac{1}{3}$$^r$$^d$ of the BSS in OCs older than 1 Gyr. This holds for NGC 2266, which is 1.1 Gyr old. 

For NGC 2266, two blue straggler stars were listed by a paper \citep{bib112}. Using \citep{bib8}, three blue stragglers were observed among the members of NGC 2266, as shown in Figure \ref{fig15}. One of these matched with Cantat-Gaudin's members and had a membership probability above 70 percent, as seen in Figure \ref{fig:10}. The NGC 2324 has one blue straggler discovered by \citep{bib105}, as shown in the previous section. The details of these BSS, including their ID, memberships, radial distances, and magnitudes, are given in Table \ref{table2}.

\begin{table*}[!ht]
\caption{Identified blue straggler star (BSS) candidates in the OCs NGC 2266 and NGC 2324. The table lists the Gaia DR3 source ID, membership probability derived from the GMM method, projected radial distance from the cluster center (in arcminutes), and the Gaia G-band magnitude of each BSS candidate.} 
\centering 
\begin{tabular}{c c c c c} 
\hline \hline 
Cluster Name & ID & Membership \% & Radial distance (arcmin) & Gmag (mag)   \\ [0.1ex] 
\hline 

NGC 2266 & 33857572896318598407 & 100 & 0.289 & 12.77\\
NGC 2266 & 3385006112734421248 &  99 & 1.094 & 13.32\\
NGC 2266 & 3385734509125347072 & 100 & 0.225 & 13.62\\
NGC 2324 & 3114552571860930176 & 100 & 1.179 & 14.33\\

\hline 
\end{tabular}
\label{table2} 
\end{table*}

\section{Dynamical Study of Clusters} \label{8}

The derived luminosity and mass functions depend on both the adopted membership probability threshold and the photometric completeness of the data. As discussed in Section~3, a threshold of $P \geq 0.7$ provides an optimal balance between contamination and completeness. At the low-mass end, the observed decline in the mass function can be attributed to a combination of effects: (i) incompleteness due to increasing photometric uncertainties and detection limits at faint magnitudes, and (ii) genuine dynamical evolution, such as the preferential loss of low-mass stars through evaporation and tidal interactions. To minimize biases, the mass function was fitted only over the magnitude range where the data remain statistically complete. Therefore, the derived slopes represent the present-day mass function rather than the initial mass function.

The luminosity function (LF) describes the distribution of stars as a function of their brightness, while the mass function (MF) represents the distribution in terms of stellar mass. These two functions are intrinsically connected through the mass--luminosity relation, which provides a direct mapping between stellar luminosity and mass. Using this relation, the observed luminosities can be transformed into stellar masses, allowing us to construct the present-day mass function (PDMF) of the clusters \citep{bib201,bib86}. Since both the LF and MF critically depend on the inclusion of genuine cluster members and the exclusion of field-star contamination, accurate membership determination is essential. In the following subsections, we present the derivation of the LF and MF for NGC~2266 and NGC~2324.

\subsection{Luminosity Function (LF)} 

\begin{figure}[h!]
  \begin{minipage}[b]{0.42\textwidth}
    \includegraphics[width=\textwidth]{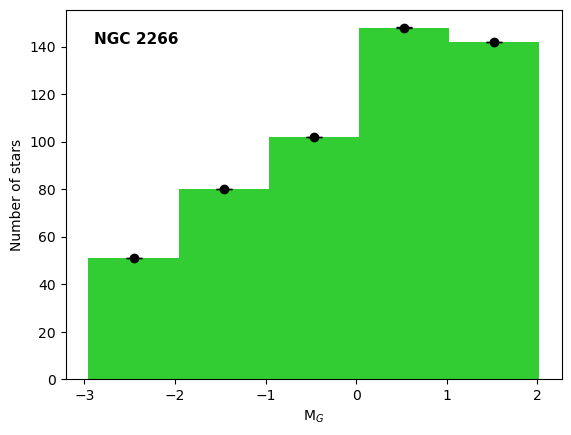}
  \end{minipage}
  \hfill
  \begin{minipage}[b]{0.42\textwidth}
    \includegraphics[width=\textwidth]{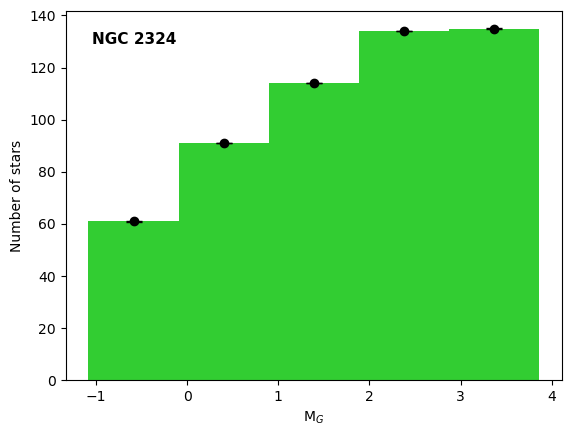}
  \end{minipage}
    \caption{ Luminosity functions of NGC 2266 (top) and NGC 2324 (bottom).
The green histograms represent the number of cluster members in each
absolute magnitude bin ($M_G$). The black points indicate the statistical
uncertainties in the star counts, estimated using Poisson statistics,
where the uncertainty in each bin is $\sigma = \sqrt{N}$ and $N$ is the
number of stars in that magnitude bin. The horizontal uncertainties
correspond to the adopted magnitude bin width.}
    \label{fig16}
\end{figure}

LF of an open cluster describes how main-sequence members are distributed across absolute magnitude intervals \citep{bib115}. Only stars brighter than the completeness limit of 19 mag were used to construct the luminosity function to avoid bias due to incompleteness at fainter magnitudes. The distance modules from subsection \ref{sec7.3} convert G magnitudes of main-sequence stars to absolute magnitudes \citep{bib201}.
We used apparent magnitudes, isochrone-based distances, and excess color values to compute absolute magnitudes ($M_G$) for the stars. These values were used to build the LF histograms for both clusters. The absolute magnitude ranges are -2.95 $\leq$ M$_{G}$ $\leq$ 2.02 for NGC~2266 and -1.08 $\leq$ M$_{G}$ $\leq$ 3.86 for NGC~2324. The mean extinction values in the $G$ band ($A_G$) for stars with membership probabilities above 70\% are 0.19 mag for NGC~2266 and 0.22 mag for NGC~2324. Figure~\ref{fig16} shows the LF histograms, each with a one mag bin width. The LF peaks near $M_G = 4$ mag for NGC~2266 and $M_G = 5$ mag for NGC 2324. The figure highlights numerous low-mass stars, indicated by the positive absolute magnitude range, which the next section discusses in more detail. We also examined the sensitivity of the luminosity function to the adopted membership-probability threshold. Tests with thresholds of 60\%, 80\%, and 90$\%$ indicate that lower thresholds introduce significant contamination from field stars, while higher thresholds reduce contamination at the expense of completeness at the faint end. The adopted threshold of $P \geq 0.7$ provides an optimal balance between completeness and purity for constructing a reliable luminosity function.

\subsection{Mass Function}

The mass function was derived from high-probability cluster members ($P \geq 0.7$) brighter than $G=19$ mag, which corresponds to the adopted completeness limit. The PDMF provides important information about the stellar content and dynamical evolution of a cluster \citep{bib67,bib68,bib30,bib66}. Since stellar masses cannot be measured directly for most cluster members, the mass distribution was obtained by converting the luminosity function into a mass function using the mass--luminosity relation derived from the best-fitting PARSEC isochrones \citep{bib61}.

The mass function was constructed by transforming absolute-magnitude bins into corresponding mass bins and fitting a power-law relation of the form

\begin{center}
log $\frac{dN}{dM}$ = $-(1+x)\times$ log$(M)$ + constant,
\end{center}

where $dN$ is the number of stars in the mass interval $dM$, $M$ is the stellar mass, and $x$ is the mass-function slope. The slope was determined through a linear least-squares fit to the logarithmic mass distribution over the mass range unaffected by incompleteness.

To assess the influence of the adopted membership criterion, we repeated the analysis using several probability thresholds ($P \geq 0.5$, 0.6, 0.7, and 0.8). The resulting mass-function slopes varied only within the uncertainties, indicating that the derived PDMF is not strongly sensitive to the exact probability threshold. Lower thresholds increase field-star contamination, whereas higher thresholds significantly reduce the number of faint, low-mass members. Therefore, $P \geq 0.7$ was adopted as an optimal compromise between completeness and membership reliability.

\begin{figure}[h!]
  \begin{minipage}[b]{0.42\textwidth}
    \includegraphics[width=\textwidth]{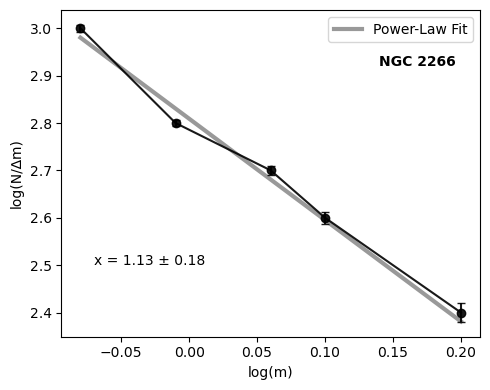}
  \end{minipage}
  \hfill
  \begin{minipage}[b]{0.42\textwidth}
    \includegraphics[width=\textwidth]{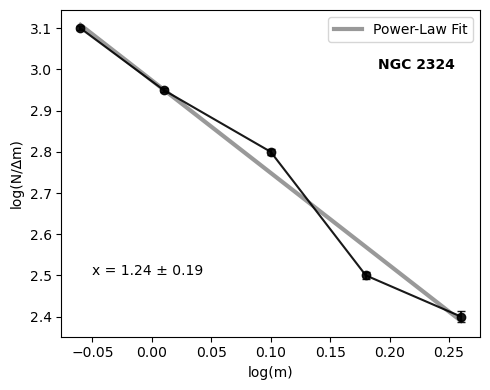}
  \end{minipage}
  \caption{ Observed stellar mass functions for the OCs NGC~2266 (top panel) and NGC~2324 (bottom panel). The vertical axis shows $\log(dN/dM)$, where $dN$ is the number of stars in a mass bin $dM$, and the horizontal axis represents $\log(M/M_{\odot})$. Only probable cluster members within the adopted completeness limit ($G \leq 19$ mag) were used in the analysis. The mass range considered for the fit is $0.75\,M_{\odot}$ to $2.0\,M_{\odot}$ for both clusters. Black circles denote the observed star counts in each mass bin, while the solid grey line represents the best-fitting power-law mass function. The derived mass function slopes are $x = 1.13 \pm 0.18$ for NGC~2266 and $x = 1.24 \pm 0.19$ for NGC~2324. Error bars correspond to Poisson uncertainties in the number counts, defined as $\sqrt{N}$ in each mass bin.}
  \label{fig17}
\end{figure}

The derived mass ranges are 0.75--1.81 M${\odot}$ for NGC~2266 and 0.80--2.00 M${\odot}$ for NGC~2324. These ranges fall within the intermediate-mass regime, for which a Kroupa-like power-law description is appropriate \citep{bib63}. The corresponding mass-function slopes are {\bf $x = 1.13 \pm 0.18$} for NGC~2266 and {\bf $x = 1.24 \pm 0.19$} for NGC~2324 (Table~\ref{table3}). These values are shallower than the canonical Kroupa IMF slope, suggesting a relative depletion of low-mass stars. The luminosity and mass functions are presented in Figures~\ref{fig16} and \ref{fig17}, respectively.

The decline in the lowest-mass bins is likely influenced by both observational incompleteness and dynamical evolution. While the finite magnitude limit of Gaia DR3 reduces the detectability of the faintest members, dynamical processes such as mass segregation and preferential evaporation of low-mass stars may further contribute to the observed flattening of the mass function. The derived total stellar masses are $752.91,M_{\odot}$ for NGC~2266 and $1285.36,M_{\odot}$ for NGC~2324, while the corresponding mean stellar masses are 1.43 and 1.79 $M_{\odot}$, respectively.

\begin{table*}[ht]
\caption{Derived mass-function parameters for the OCs NGC~2266 and NGC~2324. The table lists the stellar mass range used for the mass-function analysis ($M_{\odot}$), the slope of the present-day mass function ($x$), the total stellar mass of the cluster ($M_{\odot}$), and the mean stellar mass ($M_{\odot}$). The uncertainties in the mass-function slope are obtained from the linear fit to the logarithmic mass distribution.} 
\centering 
\vspace{0.15cm}
\begin{tabular}{c c c c c} 
\hline \hline 
Cluster & Mass range & MF slope (x) &  Total Mass & Mean Mass  \\ [0.1ex] 
\hline 

NGC 2266 & 0.75-1.81 & 1.13 $\pm$ 0.18  & 752.91 & 1.43 \\ 
NGC 2324 & 0.8-2.0 & 1.24 $\pm$ 0.19 & 1285.36 & 1.79 \\ [0.1ex] 
\hline 
\end{tabular}
\label{table3} 
\end{table*}

\subsection{Mass segregation}
\begin{figure}[h!]
  \begin{minipage}[b]{0.42\textwidth}
    \includegraphics[width=\textwidth]{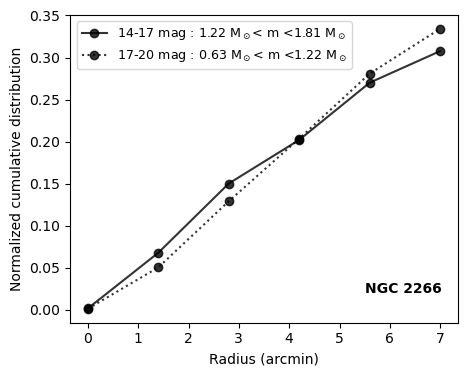}
  \end{minipage}
  \hfill
  \begin{minipage}[b]{0.42\textwidth}
    \includegraphics[width=\textwidth]{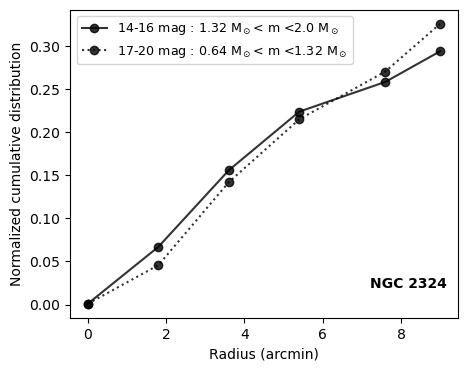}
  \end{minipage}
\caption{ Normalized cumulative radial distributions of cluster members in two different magnitude (mass) ranges for NGC 2266 (top) and NGC 2324 (bottom). The solid and dashed curves represent brighter (higher-mass) and fainter (lower-mass) stellar populations, respectively. The comparison illustrates the degree of mass segregation in the clusters, where relatively more massive stars are more centrally concentrated than lower-mass stars. The corresponding stellar mass ranges are approximately 1.22–1.81 M$_\odot$ and 0.63–1.22 M$_\odot$ for NGC 2266, and 1.32–2.0 M$_\odot$ and 0.64–1.32 M$_\odot$ for NGC 2324.}
  \label{fig:boat1}
\end{figure}

Mass Segregation is a phenomenon in which bright stars move towards the cluster center, while low-mass stars escape from the cluster into the halo region \citep{bib228,bib229,bib230,bib203,bib202,bib66,bib71,bib231,bib201}. Interactions between cluster stars transfer energy from high-mass stars to low-mass stars. As a result, bright stars appear more concentrated toward the center than low-mass stars. This effect is shown in Figure \ref{fig:boat1} and described in \citep{bib12,bib69,bib70,bib71,bib86,bib72}.

To investigate the presence of mass segregation, we examined the cumulative radial distributions of relatively bright ($G = 14$--17 mag) and faint ($G = 17$--20 mag) cluster members. Since brighter stars are, on average, more massive than fainter stars, any significant difference between their spatial distributions may indicate mass segregation within the cluster. Figure~\ref{fig:boat1} presents the cumulative radial distribution functions for both stellar groups in NGC~2266 and NGC~2324.

To quantify the differences between the two distributions, we performed a two-sample Kolmogorov--Smirnov (K--S) test. For NGC~2266, the comparison between 172 bright stars and 238 faint stars yields a K--S statistic of $D = 0.102$ and a $p$-value of 0.231, corresponding to a confidence level of 76.9\%. This result indicates that the radial distributions of the bright and faint stellar populations are statistically indistinguishable, providing no significant evidence for mass segregation in the cluster.

For NGC~2324, the K--S test was applied to 281 bright stars and 335 faint stars, yielding a K--S statistic of $D = 0.103$ and a $p$-value of 0.0768, corresponding to a confidence level of 92.3\%. Although the cumulative distribution suggests that the brighter stars are somewhat more centrally concentrated than the fainter population, the significance level remains below the commonly adopted 95\% confidence threshold. Therefore, the observed trend should be regarded as a moderate indication rather than conclusive evidence of mass segregation.

The K--S test results are summarized in Table~\ref{table4}. Overall, the analysis suggests that NGC~2324 may have undergone greater dynamical evolution than NGC~2266, resulting in a weaker mass-segregation signature. However, the statistical evidence remains modest, and the present data do not support a definitive detection of mass segregation in either cluster.

\begin{table*}[ht]
\caption{ Mass segregation analysis for the OCs NGC~2266 and NGC~2324. The table lists the stellar mass ranges (in $M_{\odot}$) used to compare the spatial distributions of higher- and lower-mass stars within each cluster. The confidence level represents the statistical significance of the mass segregation detection, derived from the Kolmogorov–Smirnov (K--S) test comparing the radial distributions of the two mass groups.} 
\centering 
\vspace{0.15cm}
\begin{tabular}{c c c c c c c} 
\hline \hline 
Cluster & Mass Range  & N$_{Bright}$ & N$_{Faint}$ & D-value & p-value & Confidence Level  \\ [0.1ex] 
\hline 

NGC 2266 & 1.81-1.22, 1.22-0.63 & 172 & 238   & 0.102 & 0.231 & 76.9$\%$  \\ 
NGC 2324 & 2.0-1.32, 1.32-0.64 & 281 & 335 & 0.103 & 0.077 & 92.3$\%$ \\ [0.1ex] 
\hline 
\end{tabular}
\vspace{1mm}
\footnotesize
\parbox{\linewidth}{
\textit{Note.} The bright and faint samples used in the K--S test were initially selected using the Gaia magnitude intervals $14 \leq G < 17$ mag and $17 \leq G \leq 20$ mag, respectively. The corresponding stellar mass ranges listed in the table were derived from the best-fitting PARSEC isochrone and are provided for reference.
}

\label{table4} 
\end{table*}

\subsection{Dynamical Parameters}
We have estimated the key dynamical parameters of the clusters under study. These include the half-mass radius, dynamical relaxation time, and dissociation time. This helps us better understand their evolutionary status and long-term dynamical stability. The half-mass radius ($R_{\mathrm{h}}$) is the radial distance that encloses half of a star cluster's total mass. To estimate $R_{\mathrm{h}}$, we used the transformation equation provided by \citep{bib73}, as adopted in \citep{bib201}. The resulting half-mass radii for NGC~2266 and NGC~2324 are 1.73~pc and 2.45~pc, respectively. The value of $R_{\mathrm{h}}$ is especially useful when considered alongside the tidal radius ($r_{\mathrm{t}}$), to investigate cluster disruption due to tidal forces \citep{bib74}. Both the internal structure and the dynamical evolution of OCs are significantly affected by tidal interactions \citep{bib75}. To study the external dynamical evolution, researchers often use the three-component Galactic model proposed in \citep{bib42}. This model helps characterize the outer regions of a cluster and determine the tidal radius $r_{\mathrm{t}}$. The tidal radius is key for studying relaxation processes and the stripping of stars by the Galactic tidal field.\\

According to \citep{bib78,bib79,bib80,bib81}, the value of Rh/Rt is related to the cluster's survival, indicating the galaxy's tidal influence on the dynamical evolution of the OCs. The Rh/Rt values obtained are 0.19 for NGC 2266 and 0.18 for NGC 2324. The lower Rh/Rt ratios indicate a more compact cluster, which supports the survival of the clusters located around smaller Galactocentric distances by compensating for their mass loss caused by the tidal field of the Milky Way Galaxy \citep{bib74}, and becomes less susceptible to tidal stripping and disruption caused by galactic gravitational forces. 

The time-scale at which the cluster will lose all traces of its initial conditions is well defined by the relaxation time T$_{R}$, which was given by \citep{bib83}:\\

~~~~~~~~~~~~~~~~~~~~~~~~~~ T$_R$ = $\frac{ 8.9 \times 10^5 \sqrt{N}  \times  R_h^{3/2} } {\sqrt{m}  \times  log(0.4N) }$ 
\\

Here, $N$ denotes the number of cluster members filtered by the GMM membership criterion with P $\geq$ 0.7, $R_{\mathrm{h}}$ is the half-mass radius, and $m$ represents the mean stellar mass of the cluster. With these parameters, the dynamical relaxation time ($T_{\mathrm{R}}$) was estimated as 7.9~Myr for NGC~2266 and 14.4~Myr for NGC~2324. As discussed in Section~\ref{sec7.3}, these values are much shorter than the ages suggested in the literature, indicating that both clusters have already had sufficient time to experience internal dynamical evolution. To formally quantify this, we compared $T_{\mathrm{R}}$ with the cluster ages by defining the dynamical evolution parameter $\tau = \mathrm{Age}/T_{\mathrm{R}}$, yielding $\tau = 153$ for NGC~2266 and $\tau = 61.7$ for NGC~2324. Since $\tau \gg 1$ in both cases, this suggests that both clusters are dynamically relaxed.\\
\\
OCs are subject to significant mass loss when they are in the stage leading to their disruption, tidally driven by the galactic field or by the transit of other molecular clouds \citep{bib85}.
The relation is given by \citep{bib84}:\\

~~~~~~~~~~~~~~~~~~~~~~~t$_{dis}$ = 250 ($\frac{ M}{300} )^{1/2}$  $\times$  ($\frac{R_h}{2}    )^{-3/2}$  \\

Here, $M$ denotes the total cluster mass and $R_{h}$ the half-mass radius. Using these parameters, the dissociation times were estimated to be approximately 492~Myr for NGC~2266 and 382~Myr for NGC~2324. These timescales are substantially longer than the corresponding dynamical relaxation times, indicating that both clusters have already attained dynamical relaxation while remaining gravitationally bound for a considerable period. The relatively long dissociation times suggest that the clusters are expected to survive for several hundred million years before being fully dispersed by the combined effects of stellar evaporation, internal dynamical evolution, and the Galactic tidal field. Although both systems are undergoing gradual mass loss, the results indicate that they are not in an advanced stage of dissolution and are likely to remain recognizable open clusters for a significant fraction of their future evolution.

\begin{table*}[!ht]
\caption{ Comparative analysis of the fundamental parameters of the OCs NGC~2266 and NGC~2324 derived in this work and reported in previous studies. The table lists astrometric, structural, and evolutionary parameters, including equatorial coordinates, proper motions, mean parallax, cluster radius, reddening, metallicity, age, heliocentric distance, Galactocentric coordinates ($X$, $Y$, $Z$), Galactocentric distance ($R_{\rm GC}$), and mass-function slope. The corresponding literature references for previously reported values are also provided.} 
\centering 
\scriptsize
\begin{tabular}{c c c c c} 
\hline \hline 
Parameters & NGC 2266  & Reference & NGC 2324 & Reference  \\ [0.5ex] 
\hline 
(R.A.) (Deg) & (100.832 $\pm$ 0.005) & {\bf This work} & (106.03 $\pm$ 0.004) & {\bf This work} \\
(DEC.) (Deg)&(26.976 $\pm$ 0.006) & &(1.047 $\pm$ 0.004) \\
&(100.829 $\pm 0.005$,) & \citep{bib220}  & (106.029, 1.045) & \citep{bib220} \\

&(100.834, 26.985) &  \citep{bib221} & (106.042, 1.038) & \citep{bib221} \\
&(100.832, 26.976) & \citep{bib43} &(106.033, 1.046)  & \citep{bib43} \\

($\mu_\alpha$ cos($\delta$)) \hspace{0.2cm}(mas/yr) & (-0.488 $\pm$ 0.021 )  & {\bf This work} & (-0.325 $\pm$ 0.005) & {\bf This work}\\ 
($\mu_\delta$) \hspace{0.1cm}(mas/yr) & (-1.247 $\pm$ 0.024) & {\bf This work} & (-0.089 $\pm$ 0.008) & {\bf This work}  \\
& (-0.76, -7.01) &\citep{bib220} & (-2.60, 1.35)& \citep{bib220}\\
& (-2.44,-4.17)&\citep{bib221} & (-2.74, 0.35)&\citep{bib221} \\
& (-0.423, -1.269) &\citep{bib44} & (-0.360, -0.049)& \citep{bib44}\\
& (-0.473, -1.272) & \citep{bib43} & -0.309, -0.094 &\citep{bib43}\\

Parallax(mas/yr) & (0.253 $\pm$ 0.010) & {\bf This work} & (0.210 $\pm$ 0.008) & {\bf This work}\\
& 0.29 & \citep{bib44} & 0.252 & \citep{bib44}\\
& 0.273 & \citep{bib43} &  0.191 & \citep{bib43}\\
Radius (arcmin) & 7.23 $\pm$ 0.47 & {\bf This work} & 10.94 $\pm$ 0.63 & {\bf This work}\\
& 3.50 & \citep{bib220} & 6.30 & \citep{bib220}\\
& 7.03 & \citep{bib56} & 8.9 & \citep{bib113}\\
Redenning & 0.17$\pm$ 0.04 & {\bf This work} & 0.22$\pm$0.06 & {\bf This work}\\
& 0.10 & \citep{bib112} & 0.25 & \citep{bib2}\\
& 0.11 & \citep{bib55} & 0.04 & \citep{bib58}\\
& 0.23 & \citep{bib221} & 0.23 & \citep{bib221}\\
& 0.17 & \citep{bib56} & 0.17 & \citep{bib204}\\
& 0.26 & \citep{bib54} & 0.26 & \citep{bib54}\\
Metallicity & 0.0084 & {\bf This work} & 0.0038 & {\bf This work}\\
& 0.007 & \citep{bib55} & 0.019 & \citep{bib60}\\
& 0.004 & \citep{bib57} & 0.004 & \citep{bib2}\\
& 0.007 & \citep{bib54} & 0.012 & \citep{bib54}\\ 
Age (log) & 9.05 & {\bf This work} & 8.9 & {\bf This work}\\
& 8.9 & \citep{bib207}   & 8.8 & \citep{bib204}\\
& 9.08 & \citep{bib55}   & 8.65 & \citep{bib2}\\
& 8.94 & \citep{bib205} & 8.83 & \citep{bib205}\\
& 9.265 & \citep{bib221} & 8.68 & \citep{bib221}\\
& 9  &  \citep{bib208}       & 8.8 & \citep{bib206}\\
& 9.15 & \citep{bib54} & 8.7 & \citep{bib54}\\
& 9.02 & \citep{bib209} & 8.65 & \citep{bib113}\\

Distance(pc) & 3550 $\pm$ 0.23 & {\bf This work} & 4180 $\pm$ 0.24 & {\bf This work}\\
& 3758 & \citep{bib207} & 2100 & \citep{bib215} \\
& 2800 & \citep{bib55} &  4169 & \citep{bib204}\\
& 2855 & \citep{bib208} & 3800 & \citep{bib2}\\
& 3311 & \citep{bib221} & 3842 & \citep{bib221}\\
& 2860 & \citep{bib54} & 4320 & \citep{bib54}\\
& 3251 & \citep{bib43} & 4214 & \citep{bib43}\\
& 3018 & \citep{bib114} & 3853 & \citep{bib114}\\
& 3114 & \citep{bib209} & - & -\\
& 3354 & \citep{bib216} & 3897 & \citep{bib216}\\
X (pc) & -3460 $\pm$ 224 & {\bf This work} & -3480 $\pm$ 200 & {\bf This work}\\
& -3169	& \citep{bib43} & -3510 & \citep{bib43}\\	
Y (pc) & -475 $\pm$ 31 & {\bf This work} & -2300 $\pm$ 132 & {\bf This work}\\
& -433 & \citep{bib43} & -2319 & \citep{bib43}\\		
Z (pc) & 631.1 $\pm$ 41 & {\bf This work} & 240 $\pm$ 14 & {\bf This work}\\
& 635	& \citep{bib43} & 242 & \citep{bib43}\\
R$_{GC}$ (pc) & 11950 & {\bf This work} & 12210 & {\bf This work} \\
& 11517 & \citep{bib43} & 12075 & \citep{bib43}\\

Mass Slope & 1.13 $\pm$ 0.18 & {\bf This work} & 1.24 $\pm$ 0.19 & {\bf This work}\\
& 2.68 & \citep{bib56} & - & - \\[0.5ex]
\hline 

\end{tabular}
\label{table5} 
\end{table*}

\section{Conclusions} \label{9}

Using \textit{Gaia} DR3 astrometry supplemented with 2MASS and LAMOST data, we performed a detailed photometric and dynamical analysis of the intermediate-age OCs NGC~2266 and NGC~2324. A total of 719 and 852 high-probability members ($P \geq 0.7$) were identified using probabilistic clustering methods, enabling improved determination of the clusters' structural and dynamical parameters. A comparison with previous studies is presented in Table~\ref{table5}. The main results of this study are summarized below.

\begin{itemize}

\item The fundamental parameters of the clusters were determined using Gaia DR3 data. The estimated ages of NGC~2266 and NGC~2324 are 1.1 $\pm$ 0.1 Gyr and 790 $\pm$ 150 Myr, respectively. The adopted heliocentric distances, based on Bayesian estimates from Gaia parallaxes, are 3.55 $\pm$ 0.23 kpc for NGC~2266 and 4.18 $\pm$ 0.24 kpc for NGC~2324. The derived metallicities are $Z = 0.0084$ for NGC~2266 and $Z = 0.0038$ for NGC~2324, while the reddening values are $E(B-V)=0.17\pm0.04$ and $E(B-V)= 0.22\pm0.06$, respectively.

\item The present-day mass functions were derived within the completeness limit using stars in the mass ranges 0.75--1.81 $M_\odot$ for NGC~2266 and 0.8--2.0 $M_\odot$ for NGC~2324. The resulting slopes are $x = 1.13 \pm 0.18$ and $x = 1.24 \pm 0.19$, indicating a relative deficiency of low-mass stars that is consistent with dynamical evolution in both clusters.

\item The estimated relaxation times are 7.9 Myr for NGC~2266 and 14.4 Myr for NGC~2324, both significantly shorter than their respective cluster ages, indicating that the clusters are dynamically relaxed. The mass-segregation analysis based on the two-sample Kolmogorov--Smirnov test reveals no statistically significant evidence for mass segregation in NGC~2266 ($D = 0.102$, $p = 0.231$), whereas NGC~2324 exhibits a mild tendency toward mass segregation ($D = 0.103$, $p = 0.0768$; confidence level $\sim92.3\%$), although the significance remains below the conventional 95\% threshold. The derived structural parameters include tidal radii of 9.13 $\pm$ 1.44 pc for NGC~2266 and 13.34 $\pm$ 0.68 pc for NGC~2324. The corresponding ratios $R_h/R_t = 0.19$ and $0.18$ indicate relatively compact cluster structures and are consistent with the clusters being dynamically evolved, although the evidence for mass segregation is weak in NGC~2266 and only mild in NGC~2324.

\item We also identify several blue straggler candidates in the clusters, including one object consistent with previous studies. Their presence provides insight into stellar interactions and binary evolution processes within the clusters.

\item The estimated dissociation times of approximately 492 Myr for NGC~2266 and 382 Myr for NGC~2324 are substantially longer than the corresponding dynamical relaxation times, indicating that both clusters have already attained dynamical equilibrium while remaining gravitationally bound systems. These results suggest that, although the clusters are subject to gradual mass loss through stellar evaporation and Galactic tidal interactions, they are expected to survive for several hundred million years before complete dissolution.

\end{itemize}

Overall, this study demonstrates the capability of \textit{Gaia} DR3 data, combined with complementary surveys, to refine the structural and dynamical parameters of OCs by improving membership determination and providing deeper photometric coverage.

\section{Acknowledgment} We thank the anonymous referees for their careful reading of the manuscript and for their constructive comments and valuable suggestions, which helped improve the clarity and presentation of this work. We express our sincere gratitude to the European Space Agency (ESA) and the \textit{Gaia} Data Processing and Analysis Consortium (DPAC) for providing \textit{Gaia} mission data, which made this research possible. We also thank the teams behind the VizieR catalog access tool (https://vizier.cds.unistra.fr), the Two Micron All Sky Survey (2MASS, https://old.ipac.caltech.edu/2mass), the LAMOST DR7 catalog (http://dr7.lamost.org), and the ESO Digitized Sky Survey (DSS, https://archive.eso.org/dss/dss). The use of software: TOPCAT (http://www.star.bris.ac.uk/~mbt/topcat/), SAO DS9 (https://sites.google.com/cfa.harvard.edu/saoimageds9), and WEBDA (https://webda.physics.muni.cz) was instrumental in our analysis, and we are grateful to the developers of \texttt{pyUPMASK} (https://github.com/msolpera/\texttt{pyUPMASK}) and GMM (https://github.com/Ransaka/GMM-from-scratch) for their clustering techniques, which significantly aided our calculations of membership probabilities. We thank Mr. Kuldeep Belwal for the fruitful scientific discussions that greatly contributed to this work. We also acknowledge constructive feedback from our colleagues, mentors, and anonymous reviewers, whose contributions improved the quality of this work. 

\nolinenumbers
\makeatletter
\def\@biblabel#1{}
\makeatother

\bibliographystyle{jasr-model5-names}
\biboptions{authoryear}
\bibliography{References1}

\end{document}